\documentclass{aa}  
\usepackage{graphicx}
\usepackage{url}

\usepackage{natbib}
\bibpunct{(}{)}{;}{a}{}{,} 

\usepackage{txfonts}
\def\micron{\hbox{$\mathrm{\mu m}$}}
\def\laboca{\hbox{LABoCa}}
\def\position{\hbox{$\alpha = 21^h42^m43\fs7$, $\delta =
    -44\degr25\arcmin30\farcs0$ (J2000)}}

\begin{document}

\title{Submillimeter observations of the J2142-4423 Ly$\alpha$ protocluster at
  $z = 2.38$\thanks{This work is based on observations made with the APEX
    Telescope. APEX is a collaboration between the Max-Planck-Institut f\"ur
    Radioastronomie, the European Southern Observatory, and the Onsala Space
    Observatory.}}
\subtitle{}

\titlerunning{\laboca\ observations of the J2142-4423 protocluster}

\author{A.~Beelen\inst{1}
  \and
  A.~Omont\inst{2}
  \and
  N.~Bavouzet\inst{1}
  \and
  A.~Kov\'{a}cs\inst{3}
  \and
  G.~Lagache\inst{1}
  \and
  C.~De Breuck\inst{4}
  \and
  A.~Weiss\inst{3}
  \and
  K.~M.~Menten\inst{3}
  \and
  J.~W.~Colbert\inst{5}
  \and
  H.~Dole\inst{1}
  \and
  G.~Siringo\inst{3}
  \and
  E.~Kreysa\inst{3}
}

\authorrunning{Beelen et al.}

   \offprints{A. Beelen}

   \institute{Institut d'Astrophysique Spatiale,
     b\^at 121 - Universit\'e Paris-Sud, 
     91405 Orsay Cedex / France \\
     \email{alexandre.beelen@ias.u-psud.fr}
     \and
     Institut d'Astrophysique de Paris,
     CNRS and Universit\'e Pierre et Marie Curie,
     98bis, bd Arago, 
     75014 Paris /  France 
     \and
     Max-Planck-Institut f\"ur Radioastronomie,
     Auf dem H\"ugel 69, 
     53121 Bonn / Germany
     \and
     European Southen Observatory, Karl Schwarzschild Stra{\ss}e 2,
     D-85748 Garching / Germany
     \and
     \textit{Spitzer} Science Center, 
     California Institute of Technology, 
     Mail Code 220-6, Pasadena, 
     CA 91125-0600 / USA}

     \date{Received / Accepted}
     
 
\abstract
  {}
  {We present observations aimed at exploring both the nature of Ly$\alpha$
    emitting nebulae (``Ly$\alpha$ blobs'') at $z=2.38$ and the way they
    trace large scale structure (LSS), by exploring their proximity to
    ``maximum starbursts'' through submillimeter emission.  Our most
    important objectives are to make a census of associated submillimeter
    galaxies (SMGs), check their properties, and look for a possible
    overdensity in the protocluster J2142-4426 at $z=2.38$.}
  {We used the newly commissioned Large APEX Bolometer Camera (\laboca) on
    the Atacama Pathfinder EXperiment (APEX) telescope, in its Science
    Verification phase, to carry out a deep $10\arcmin\times10\arcmin$ map
    at 870~\micron, and we performed multiple checks of the quality of data
    processing and source extraction.}
  {Our map, the first published deep image, confirms the capabilities of
    APEX/\laboca\ as the most efficient current equipment for wide and deep
    submm mapping. Twenty-two sources were securely extracted with
    870~\micron\ flux densities in the range $3-21\,\mathrm{mJy}$, rms noise
    $\sim 0.8-2.4\,\mathrm{mJy}$, and far-IR luminosities probably in the
    range $\sim 5-20\times 10^{12} \, \mathrm{L_\odot}$. Only one of the
    four 50~kpc-extended Ly$\alpha$ blobs has a secure 870~\micron\
    counterpart. The 870~\micron\ source counts in the whole area are
    marginally higher than in the SHADES SCUBA survey, with a possible
    over-density around this blob. The majority of the $3.6-24~\micron$ SEDs
    of the submillimeter sources indicate they are starburst dominated, with
    redshifts mostly~$\gtrsim 2$. However, there is evidence of a high-z AGN
    in $\sim30\%$ of the sources.}
  {}


   \maketitle

\section{Introduction}

Narrow-band surveys are a very powerful technique for detecting Lyman
$\alpha$ emission from various structures at high redshift. In addition to
numerous Ly$\alpha$ galaxies, they have revealed the existence of very
extended ($30-200\,\mathrm{kpc}$) Ly$\alpha$ nebulae, generally called
Ly$\alpha$ blobs \citep[see e.g.][ and references
therein]{Francis1996,Steidel2000,Palunas2004,Matsuda2007}. Currently, only
about 10 blobs with diameter, $\Phi, \gtrsim 60\, \mathrm{kpc}$, are known,
and twice more with $\Phi \gtrsim 50\, \mathrm{kpc}$ \citep[see e.g.][ and
references therein]{Matsuda2004,Matsuda2007,Smith2007}. They are generally
part of larger structures of Ly$\alpha$ emitters of various sizes.

The origin of the Ly$\alpha$ emission in such blobs is still a matter of
discussion. Widespread Ly$\alpha$ emission is known to be associated with
starbursts and shocks. Multiple supernovae explosions, including galactic
outflows, associated with giant starbursts, could be excellent candidates
for powering Ly$\alpha$ emission. But accretion cooling flows
\citep[e.g.][]{Nilsson2006, Smith2007} and an obscured AGN have also been
suggested as other possible power sources. Ly$\alpha$ blobs present many
similarities with high redshift radio galaxies, which often have giant
Ly$\alpha$ haloes up to $150\,\mathrm{kpc}$ \citep[see][ for a
review]{Miley2008}. However, all Ly$\alpha$ blobs observed to date remain
undetected in deep radio observations, excluding the possibility that their
Ly$\alpha$ emission is powered by a radio-loud AGN like in the radio
galaxies.Whatever is their power source, there is evidence that Ly$\alpha$
blobs are strong infrared emitters. For instance, most of the 35 Ly$\alpha$
blobs with $\Phi \gtrsim 30\,\mathrm{kpc}$ identified by \citet{Matsuda2004}
in the SA~22 region have been detected in deep Spitzer IRAC/MIPS
observations \citep[][ Huang et al. in preparation]{Yamada2007}.

The 110~Mpc filament with 37 Ly$\alpha$-emitting objects around the galaxy
protocluster J2143-4423 at $z=2.38$ is one of the largest known structures
at high $z$ \citep{Palunas2004, Francis2004}. In addition to its compact
Ly$\alpha$ galaxies, it also includes four extended Ly$\alpha$ blobs ($\Phi
\gtrsim 50\, \mathrm{kpc}$). From Spitzer/MIPS observations,
\citet{Colbert2006} have reported the detection of a number of 24~\micron\
sources associated with Ly$\alpha$ emitters. In a central
$8\arcmin\times10.5\arcmin$ area of the J2143-4423 region, they have
detected five 24~\micron\ sources with a $24\,\micron$ flux density, $F_{24}
\gtrsim 150\ \mathrm{\mu Jy}$, closely associated with Ly$\alpha$ emitters,
three of them being extended Ly$\alpha$ blobs. The far-infrared
luminosities, L$_{FIR}$, that they inferred, range from $0.5$ to $5\,
10^{13}\, \mathrm{L_\odot}$. There are also three other similar 24~\micron\
sources with a looser (within $\sim 10\arcsec$) association to the
50~kpc-Ly$\alpha$ blobs.  \footnote{The nomenclature about Ly$\alpha$ blobs
  is still a bit confused. A Ly$\alpha$ blob (or LAB) refers clearly to an
  extended or resolved region of Ly$\alpha$ emission. However, the evidence
  for extension depends on the quality of the optical image. In the
  prototype region SA 22, there is a list of several tens of LABs down to an
  extension of $\sim 5\arcsec$ \citep[see e.g. Table 1 of][]{Geach2005}. In
  the J2143-4423 region, extensions are published only to $\sim 7\arcsec$
  ($\sim50\, \mathrm{kpc}$), for four blobs, B1, B5, B6, B7 \citep[see Table
  4 of][]{Palunas2004}. However, other unresolved Ly$\alpha$ emitters are
  also labeled B2, B4, B8 and B9 \citep{Francis1996, Francis1997,
    Palunas2004, Colbert2006}. To avoid confusion, we will call extended
  objects of the first group 50~kpc-Ly$\alpha$ blobs}

Such a concentration of Ly$\alpha$ blobs suggests an exceptionally large
structure (110~Mpc) above $z=2$. However, neither Ly$\alpha$, nor
24~\micron\ emissions can give an unambiguous answer to the exact nature of
the galaxies and their star formation properties. Submillimeter observations
are fundamental to determine the far-IR luminosity of the various
24~\micron\ sources associated with Ly$\alpha$ emitters, and thus their star
formation rate. The SCUBA 870~\micron\ study by \citet{Geach2005} of the
similar structure SA~22 discovered by \citet{Steidel2000}, with similar
Ly$\alpha$ luminosity and a larger number of Ly$\alpha$ blobs
\citep{Matsuda2004}, detected about 20\% of the blobs (with possible
statistical detection of the full sample). This could indicate that their
FIR luminosity FIR luminosity is starburst powered, with FIR luminosities in
the ultra-luminous regime ($> 5\, 10^{12}\ \mathrm{L_\odot}$), equivalent to
a star formation rate approaching $10^3\, \mathrm{M_\odot/yr}$. However, the
relation between the Ly$\alpha$ and even 24~\micron\ emission of the blobs
to their star formation rate is probably not straightforward, as shown by
the surprising non-detection of the strongest ($17\, \mathrm{mJy}$) SA~22
SCUBA source in the high-resolution submillimeter imaging of
\citet{Matsuda2007}, indicating extended submm emission.
   
In this context, we used the newly commissioned Large Apex Bolometer Camera
(\laboca) \citep[ 2008 in prep.]{Siringo2007} to check the properties of
ultra-luminous starbursts in the J2143-4423 Ly$\alpha$ blobs and their
surrounding field. The paper is organized as follows~: Section~\ref{sec:obs}
describes the observation with \laboca\ of the central part of the z=2.38
Ly$\alpha$-emitter over-density at \position, which was observed at
24~\micron\ by \citet{Colbert2006}, Section~\ref{sec:analysis} presents our
analysis and in Section~\ref{sec:final} we discuss the results and report
our conclusions.  Throughout the paper, we assume a concordance
$\Lambda$-cosmology with $H_0=71\, \mathrm{km\ s^{-1}\ Mpc^{-1}}$,
$\Omega_\Lambda=0.73$ and $\Omega_m=0.27$.

\section{Observation}
\label{sec:obs}

\subsection{\laboca\ Observations}

\begin{figure}
  \resizebox{\hsize}{!}{\includegraphics{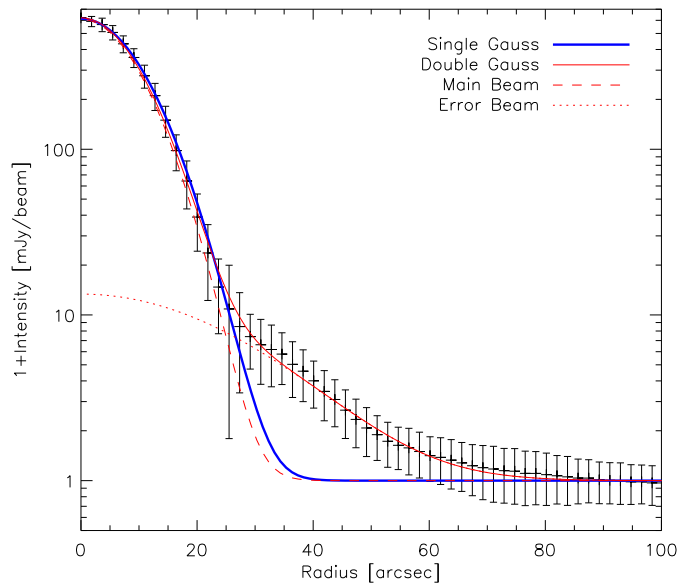}}
  \caption{Effective radial beam profile of \laboca, combining all the
    bolometers that were usable during the observing run. The error bars are
    derived from the standard deviation of the pixels in the map. The
    profile is well described either by a single Gaussian of
    HPBW=$20.4\pm0.5$\arcsec\ (thick line), or by a Gaussian main beam of
    HPBW=$19.1\pm0.6$\arcsec\ (dashed line) and a Gaussian error beam of
    HPBW=$54\pm5$\arcsec\ (dotted line), once deconvolved from the median
    Mars diameter at the time of observation.  The peak ratio between the
    error and the main beam is 2.1\%.  }
 \label{fig:beam}
\end{figure}

Observations were conducted using \laboca\ \citep[ 2008 in
prep.]{Siringo2007} installed on the Atacama Pathfinder EXperiment
\citep[APEX,][]{Gusten2006}. \laboca\ is an array consisting of 295
bolometers arranged in 9 concentric hexagons, operating in total power at
280~mK, with a half-power spectral bandwidth from 313 to 372~GHz, and an
effective frequency of 345~GHz~(870~\micron). The number of bolometers with
sky response is 266, of which 15 show signature of cross talk and 18 have
very low sensitivity; in total 33 bolometers have been discarded from the
data analysis. Two additional bolometers have been blinded in order to
record the temperature variation of the detector wafer.  The complete array
field of view covers 11.4\arcmin.
Using fully sampled observations of Mars, we derived, for each detector, the
relative gains and relative positions of the bolometers; the latter were
found to be stable within 1\arcsec. The gains have been normalized using the
median value of all valid bolometers. The effective radial beam profile
deduced from Mars observations is shown in Fig.~\ref{fig:beam}. Once
deconvolved from the size of Mars, varying from 6.5\arcsec\ to 6.9\arcsec
during the observations, the beam profile can be approximated, within a few
percent, by a single Gaussian beam of half power beam width (HPBW) of
$20.4\pm0.5\arcsec$ for point source studies.

During the two disjoint observing periods, 2007 July 07-15 and August 15-28
(ESO program ID : 078.F-9030(A) and MPIfR 080.F-9502(A), respectively), the
atmospheric conditions were excellent, with typical zenith opacities between
0.07 and 0.17 and a median value of 0.12 at 870~\micron.
The telescope pointing was checked once an hour using the nearby radio
source \object{PMN~J1957-3845} and was found to be stable within a few arc
seconds in azimuth and elevation.
The focus setting in the Z direction was checked regularly every 2 to 3
hours on Mars, and at least every day in the X and Y direction and was found
to be stable in each direction.
The target area was mapped with a raster of scans in horizontal coordinates
tracing spirals displaced from each other in a way to obtain optimum spatial
sampling. Simulations were made using \textsc{BoA} (see subsection below) in
order to derive the best spiral parameters to obtain a fully sampled map of
the sky with the actual foot pattern of \laboca\ and with little overhead
compared to classical On-The-Fly observing mode. The total on sky
integration time was 14+7=21~hours.

The absolute flux calibration is based on observations of Mars during the
first observing run and a model\footnote{see
  \url{ http://www.lesia.obspm.fr/~lellouch/mars/}} developed by
E.~Lellouch, H.~Amri and R.~Moreno, using general climate model predictions
of martian surface and sub-surface temperatures \footnote{see
  \url{ http://www-mars.lmd.jussieu.fr/}}, and radiative transfer within the
surface. The predicted total flux of Mars varied from 660 to 711~Jy, over
the first observing period. After correction for the angular extent of Mars,
the derived calibration factor for \laboca\ is $6.8\pm0.5\,\mathrm{Jy/\mu
  V}$. The absolute flux calibration uncertainty is thus about 7\%. Flux
densities quoted in the following do not include this uncertainty since it
is only useful when compared to other instruments.

\subsection{Data Reduction}

The data were reduced with an updated version of the BOlometer Array
Analysis Software (\textsc{BoA}), a newly designed free software package to
handle bolometer array data. \textsc{BoA} is a collaborative effort of
scientists at the Max-Planck-Institut f\"ur Radioastronomie (MPIfR),
Argelander-Institut f\"ur Astronomy (AIfA), Astronomisches Institut Ruhr
Universit\"at Bochum (AIRUB) and Institut d'Astrophysique Spatiale (IAS),
with the primary goal of handling data from \laboca\ at APEX, but it can
also be used to process data acquired with other instruments such as ASZCa
\citep{Dobbs2006} and (p)ArT\'eMiS at APEX or MAMBO \citet{Kreysa1998} at
the IRAM 30-m telescope.

The data are corrected for atmospheric opacity at the time of the
observation by a linear interpolation from a combination of skydips and APEX
radiometer measurements. Using the two blind bolometers, it is possible to
correct for the temperature variation of the helium-3 stage over the
complete observation. Flat-fielding was then applied based on bolometer
relative gains on known primary calibrators. Before any further processing,
the data stream was flagged according to the telescope pattern to avoid high
accelerations, responsible for microphonics, as well as low and high speed,
in order to properly disentangle the sky signal spatial frequencies from the
atmospheric emission.
The sky emission was iteratively estimated and removed using the median
value of all valid bolometers, over the whole array first, then by grouping
the bolometers by electronics boxes or cables to remove any remaining
correlated signal due to the electronics or micro-phonics pick-up. This
removed the atmospheric emission satisfactorily across the array and no or
little correlated signal is seen after applying this procedure. An iterative
despiking is applied to the data, before removing a linear baseline. A last
flagging was set up on the bolometers according to the Median Absolute
Deviation of their variance in order to filter very noisy or dead
bolometers.
Finally the data are weighted by the inverse of their variance, and gridded
on the sky with a pixel size of one third of the beam, about 6\arcsec, the
flux in each pixel being the weighted-average of all bolometers observing
that position, producing a signal ($S$) and a weight ($W$) map.
Each scan was visually inspected in order to remove obvious problems like
wrong sky noise removal, data corruption, canceled scans or bolometers
warming up during the observations.

Other methods were tried to remove atmospheric noise, in particular one
based on principal component analysis (PCA), where the signal is transformed
to a coordinate system where the first axis correspond to the greatest
variance of the data. By removing the first few coordinates and projecting
the data back to its original basis, the atmospheric emission, responsible
for most of the signal is then removed. Both methods, removing median signal
and the PCA, show very similar results.  Moreover the data was independently
reduced following a totally different approach \citep[see][ for an
introduction]{Kovacs2006} by using the \textit{mini-crush} program.  The
final map and source list are compatible with the result presented here.

\begin{figure}
  \resizebox{\hsize}{!}{\includegraphics{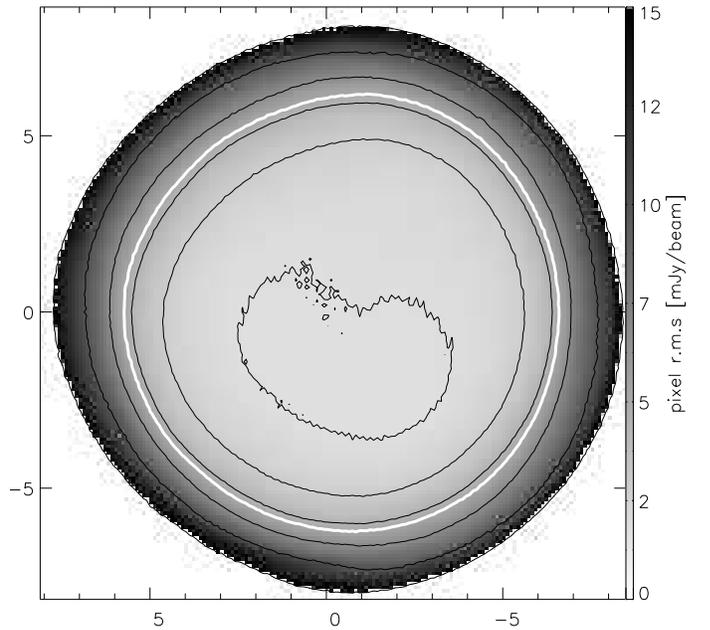}}
  \caption{Noise map around the $z=2.38$ galaxy protocluster J2143-4423.
    Contours are at pixel rms 1.9, 2.9, 4.4, 6.6, 9.9, 15.0 mJy/beam in
    exponential progression, the thick white contour corresponds to the
    region of interest with pixel rms below 5~mJy/beam (see text). The
    axes denote the offsets in arc minutes from the map center
    (\position).}
\label{fig:rms}
\end{figure}

On the best 10\% observed region, corresponding to a homogeneously observed
surface of $45\,\mathrm{arcmin}^2$, the pixel rms is 1.93~mJy/beam, which
correspond roughly to a point source sensitivity rms of 1.4~mJy, as the
pixel size is a fraction of the beam. By rescaling the weight map with this
value, we produced a noise map shown in Fig.~\ref{fig:rms}.  The noise is
fairly flat in the center of the map with a pixel rms between 1.95 and
5~mJy/beam and increases rapidly toward the edge of it. In the following we
will limit our study to the area with better than 5~mJy/beam pixel rms,
corresponding roughly to a surface of $120\,\mathrm{arcmin}^2$, or 60\% of
our map; this corresponds to the white thick line in Fig.~\ref{fig:rms}.

\subsection{Source Extraction}

\begin{figure*}
 \centering 
 \includegraphics[width=17cm]{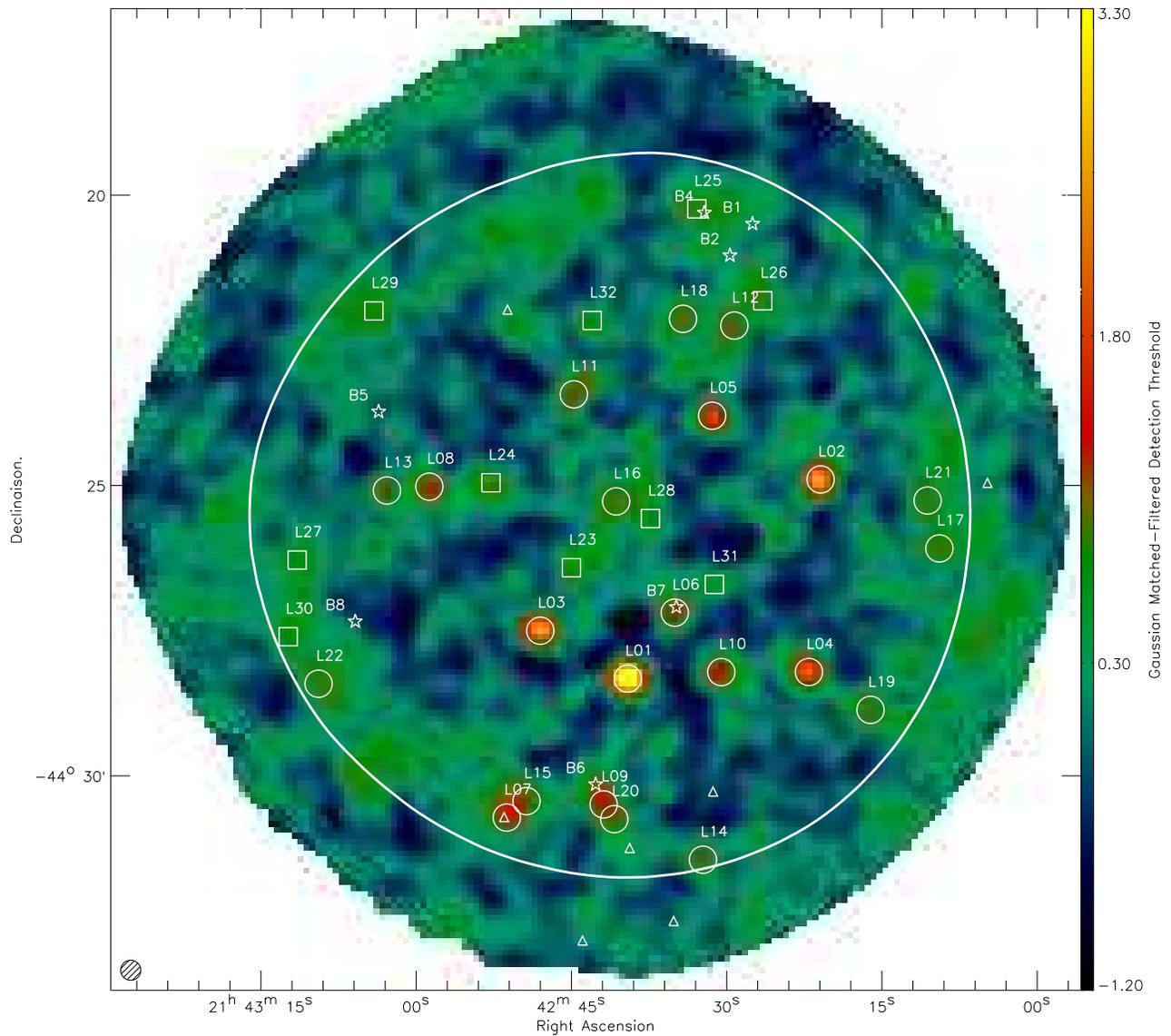}
 \caption{The 870~\micron\ \laboca\ Gaussian-matched-filtered detection
   threshold map around the $z=2.38$ galaxy protocluster J2143-4423. The map
   pixels are about $6\arcsec\times6\arcsec$.  Detected sources with
   detection threshold greater than 1.2~(1.0) are circled~(boxed). This
   detection threshold corresponds roughly to a signal-to-noise ratio of
   4.3~(3.4).  The thick white contour corresponds to the region of interest
   (see text).  The \laboca\ HPBW beam shape is represented in the lower
   left corner. Ly$\alpha$ Blobs present in the field are represented by
   open stars \citep{Palunas2004}. Triangles are QSOs detected by
   \citet{Francis2004}. }
 \label{fig:s2n}
\end{figure*}

As seen in Fig.~\ref{fig:rms}, the noise is not uniform over the observed
field, and the signal (S) and weight (W) map were gaussian-matched-filtered
to take this into account. The FWHM of the gaussian ($P$) filter was set to
$\sqrt{20.4^2+3^2}=20.7\arcsec$, in order to take into account the HPBW of
\laboca\ and a typical pointing error of 3\arcsec. Following
\citet{Serjeant2003}, the detection threshold map after this noise-weighted
convolution can be expressed as
$$ \frac{F}{\Delta F} =
\frac{(S\ W) \otimes P}{\sqrt{W \otimes P^2}}, \label{eq:filter} 
$$%
and is presented in Fig.~\ref{fig:s2n}.

\begin{figure*}
  \centering
  \includegraphics[width=17cm]{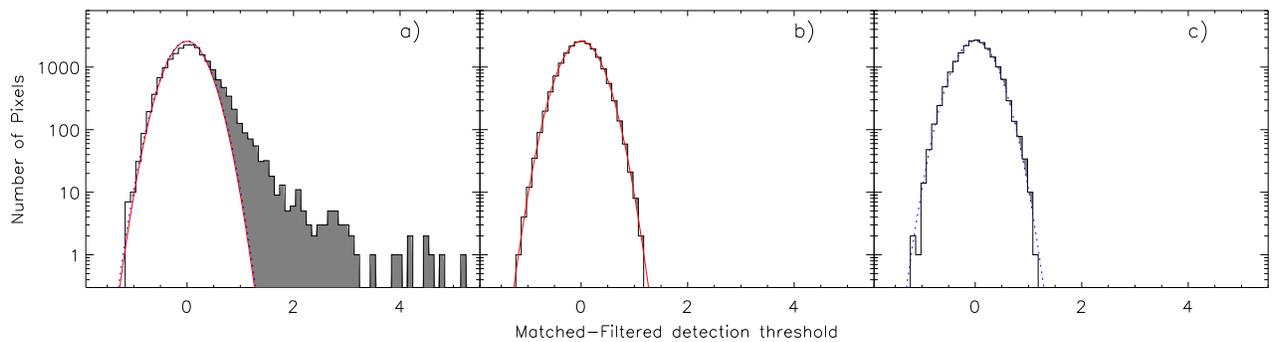}
  \caption{Histograms in Log-Normal coordinates of the pixel signal-to-noise
    values of the gaussian-machted maps. a) in the signal map, where a clear
    excess (shaded area) on the positive-side is present compared to the
    jackknife histogram fit (solid line) or the astrometry-corrupted fit
    (dashed-line). b) in the mean jackknife and c) astrometry-corrupted map
    (see text) which can both be very well fitted by a Gaussian distribution
    (solid line in b and dashed line in c). }
 \label{fig:histogram}
\end{figure*}

In order to test the robustness of the galaxy candidate identification, two
tests were performed. The first test is derived from the jackknife test
where the observations are divided into two randomly chosen samples of equal
size. The difference of the two maps should remove both resolved and
confused sources signal and its histogram should be described by a pure
Gaussian. Figure~\ref{fig:histogram} presents the average result of 100
jackknifed maps. A Gaussian distribution fits very well the negative part of
the jackknife histogram, and shows no positive-side excess indicating that
the jackknife procedure did remove all source signal.
For the second test, the relative positions of the bolometers were shuffled
inside the field of view, corrupting their astrometry. In the resulting
co-added map, any coherent source on the sky should be smeared out in the
noise, and any positive excess on the map pixel distribution will be
suppressed, while the noise properties of the resulting map remain similar
to the real map.  Figure~\ref{fig:histogram} shows the histogram of one
astrometry-corrupted map, whose negative part is very well fitted by a
Gaussian distribution and which presents no or very little excess on the
positive side, due to the smearing over the map of the sources signal.

The histogram of the detection threshold map is also shown in
Fig~\ref{fig:histogram}.  The negative part of the histogram is highly
Gaussian and is well described by the over-plotted Gaussian fits of the
jackknifed and astrometry-corrupted distribution, showing a clear excess on
the positive side. The galaxy candidates clearly account for this excess, as
the residual map, after source extraction, does not show such a significant
excess.

\begin{figure}
  \resizebox{\hsize}{!}{\includegraphics{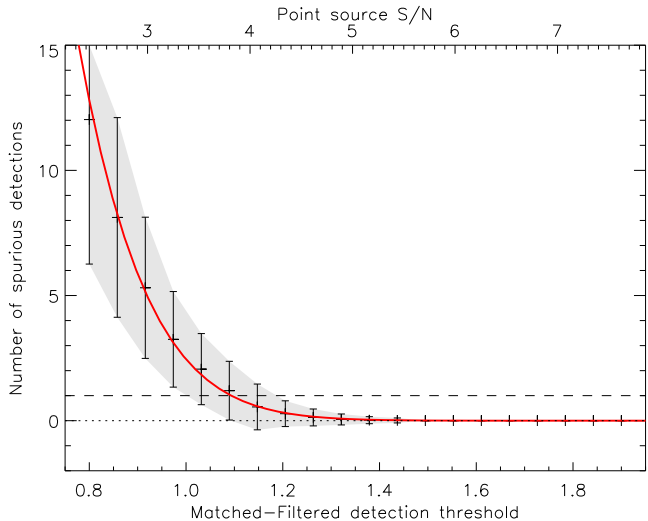}}
  \caption{Expected average spurious detection rate in our map as a function
    of the Gaussian matched-filtered signal-to-noise ratio. The solid line
    is an exponential fit to the data. Error bars are $1\,\sigma$ standard
    deviations of the jackknifed maps. The upper axis presents roughly the
    corresponding signal-to-noise ratio for point sources.}
  \label{fig:threshold}
\end{figure}

In order to determine the false detection rate, we produced 100 jackknifed
maps and performed source extraction, using a \textsc{clean} algorithm
\citep{Hogbom1974}, with different detection thresholds. This gives a
realistic picture of the number of spurious sources expected in our map at a
given threshold. However, the jackknifed maps are free of confusion noise,
therefore the number of spurious sources might be a slightly underestimated.
The result is shown in Fig.~\ref{fig:threshold}, where the solid line is an
exponential fit to the data as $F(r) = a\ \exp(-r^2/b)$. We find that a
Gaussian matched-filtered detection threshold of 1.2 results in at most one
spurious source expected at random, whereas with a threshold of 1.0 we
expect about four spurious sources.  Translated into the non-filtered map,
these thresholds correspond roughly to a signal-to-noise ratio for point
source of 4.3 and 3.4, respectively.

\begin{figure*}
  \includegraphics[width=17cm]{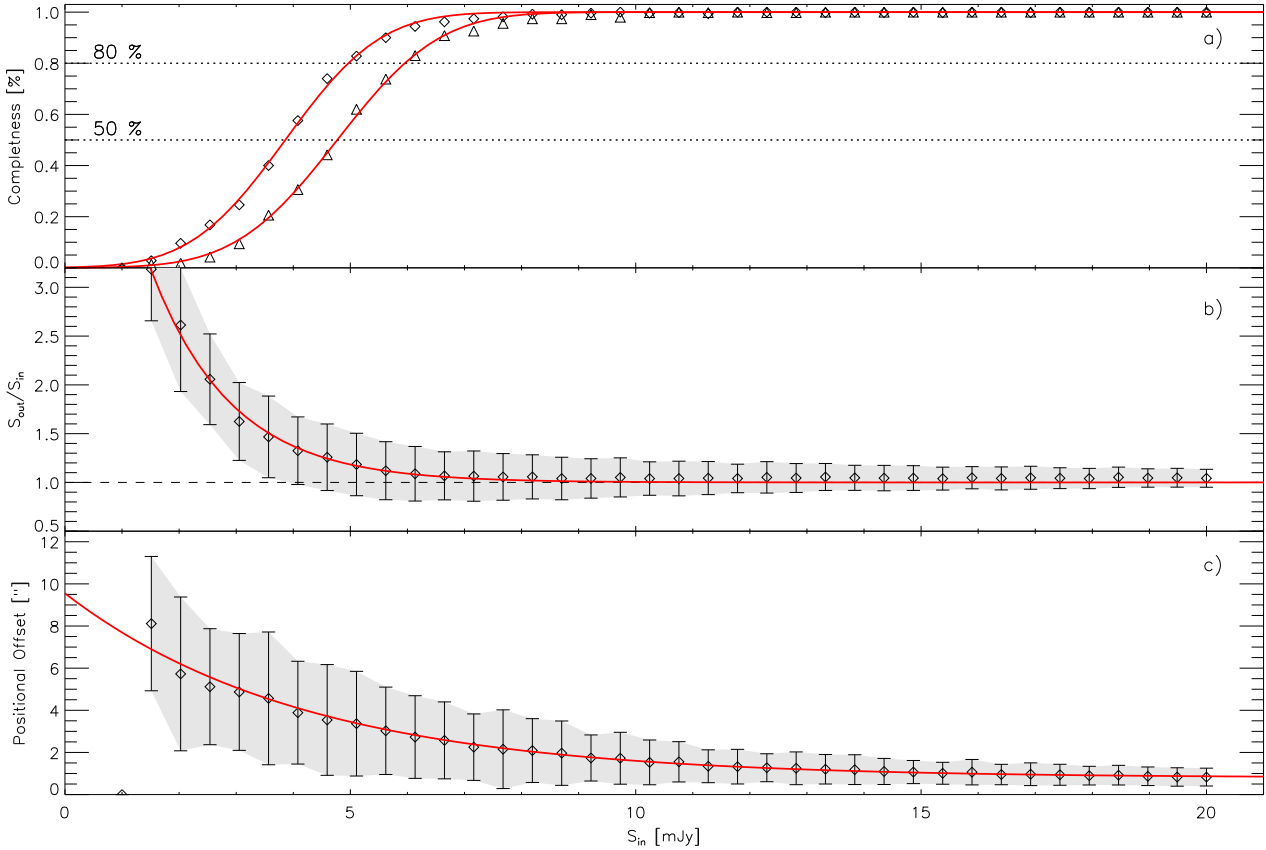}
  \caption{Results of Monte-Carlo simulations with a
    Gaussian-matched-filtered detection threshold ratio of 1.0 (diamond) and
    1.2 (triangle), to test for completeness, flux boosting and position
    uncertainties as a function of the input flux. Error bars are the
    $1\,\sigma$ standard deviation of the jackknifed maps.  a ) Completeness
    function, i.e.  the fraction of recovered sources. b) Flux boosting,
    i.e. the ratio between the extracted and input flux. c) Positional
    uncertainties, i.e.  difference between the input and recovered source
    positions. The solid line represent fits to the data, using an error
    function for a), as $f(S_{in}) = (1+\mathrm{Erf}((S_{in}-a)/b))/2$ with
    $a = 3.8\,(4.7)$ and $b = 1.8\,(2.0)$ for a detection threshold of 1.0
    (1.2) and an exponential function of the form $f(S_{in}) = a\exp(-b\,
    {S_{in}})+1$ with $a=6.3$ $b=0.7$ for b) and $f(S_{in}) = a\exp(-b\,
    {S_{in}})+c$ with $a=8.7$ $b=0.2$ and $c=0.8$ for c). Results for b) and
    c) are similar for the two detection thresholds and only one is
    presented for clarity. }
  \label{fig:montecarlo}
\end{figure*}

The measured fluxes in a map are biased toward high values due to
instrumental, atmospheric and confusion noise, this is known as the ``flux
boosting'' effect. We performed Monte-Carlo simulations to test the
completeness, flux boosting and positional uncertainties of the source
extraction algorithm. Using one jackknifed map we added, at a random
position in the map, one source with flux between 1 and 20~mJy in steps of
0.5~mJy. Repeating this process 500 times for each flux bin, source
extraction was made by selecting all regions with a detection threshold,
${F}/{\Delta F} > 1.2$, and we computed the flux densities within an
aperture of 20.4\arcsec\ directly in the non-filtered map. The photometry
correction is derived from the high signal to noise map of Mars and is about
7\%.  The results of these simulations are presented in
Fig.~\ref{fig:montecarlo} for a Gaussian matched-filtered signal-to-noise
threshold ratio of 1.00. As expected, the source extraction performs well in
extracting all the brighter sources but degrades when dealing with fainter
sources. At a flux density of 4~mJy, the completeness is about $\approx50$
per cent. The effect of flux boosting is clearly seen for faint flux
densities, where the instrumental noise tends to favor detection of sources
coinciding with positive noise peaks.  At a flux density level of 4~mJy,
this effect is on the order of 35 per cent, decreasing exponentially for
higher fluxes. Last, the positional offset due to the source extraction
algorithm is found to be approximately 4\arcsec\ for a flux density of 4~mJy
and decreasing for higher fluxes. For a flux density of 10~mJy, this
positional error is about 1\arcsec, corresponding to one sixth of a pixel,
negligible with respect to the pointing error.

Within the region of interest of the Gaussian matched-filtered map, we
performed source extraction with a threshold of ${F}/{\Delta F} > 1.0$,
leading to the detection of 22 sources with ${F}/{\Delta F} > 1.2$ and 10
additional sources below this threshold where we expect, in total on the
full sample, between 1 and at most 5 spurious sources. The source candidates
are listed in Tab.~\ref{tab:sources}, in decreasing Gaussian
matched-filtered detection threshold ($F/\Delta F$) order and are also seen
in Fig.~\ref{fig:s2n}. The listed flux densities are corrected for the
boosting flux effect and are in the range of 2.9 to 21.1~mJy. The flux
uncertainties are computed from dispersion in the sky annulus and the
uncertainty in the mean sky brightness. The quoted uncertainties are also
corrected for the boosting flux effect but do not propagate its
uncertainties.

\section{Analysis}
\label{sec:analysis}

\subsection{Number counts}
\label{sec:counts} 

\begin{figure}
  \resizebox{\hsize}{!}{\includegraphics{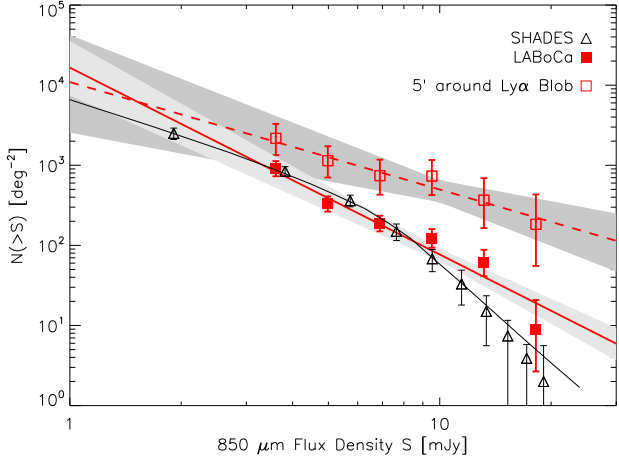}}
  \caption{Cumulative number counts at 870~\micron\ (full square) based on
    the sources detected in this paper. The errors bars are the $1\sigma$
    two-sided Poissonian confidence level. The thick plain line is a power
    law fit to the data (see text). Also shown are the 850~\micron\
    cumulative number counts and fit of the SHADES survey with triangles and
    thin line, respectively \citep{Coppin2006}. The 850~\micron\ number
    counts have been rescaled to 870~\micron\ with a spectral index of
    $\nu^2$. Furthermore, the number counts in a 5\arcmin\ diameter circle
    around the source L06 are presented in open squares and a dashed line
    that has been fitted to them is also shown. Shaded area represent the
    1~$\sigma$ uncertainties of the fit. }
  \label{fig:counts}
\end{figure}

We performed a cumulative number count analysis on the detected sources
using the fluxes corrected for the flux boosting effect and for the
completeness of the survey as discussed previously.  Based on 22 sources,
the 870~\micron\ integrated number counts are presented in
Fig.~\ref{fig:counts} and are well described by a power law of the form
$N(>S) = N_0 (S/S_0)^{-\alpha}$ : with the parameter $S_0=10\ \mathrm{mJy}$,
we derived $N_0 = 70\pm12\, \mathrm{deg^{-2}}$ and $\alpha = 1.9\pm 0.2$. On
the same figure, we also show the cumulative combined SHADES number counts
obtained by \citet{Coppin2006} on a $720\,\mathrm{arcmin}^2$ field down to
an rms of 2~mJy where $>100$ galaxies were uncovered. Our sample of 22
sources is too small to allow any meaningful detailed analysis of the number
count distribution, and, for example, we do not see any trend for a break in
the power law as in the SHADES data. However, we tentatively observe an over
density of sources at $S_{870\,\micron} > 10\, \mathrm{mJy}$. This is
confirmed when we look at the number counts of 4 sources in a 5\arcmin\
diameter circle around the source L06, identified as a Ly$\alpha$ blob (see
below), with a power law parametrized by $N_0 = 498 \pm 152\,
\mathrm{deg^{-2}}$ and $\alpha = 1.3\pm0.5$. The density of submillimeter
galaxies is then one order of magnitude, at a $3\,\sigma$ level, higher than
for unbiased submillimeter field galaxies.

\subsection{Mid-IR Source Identifications}
\label{sec:label}

Identification of millimeter and submillimeter galaxies has mainly used the
strong correlation between the far-IR and radio luminosities, tracing
respectively, warm dust heated by hot, young O-type stars and non-thermal
emission resulting from supernova explosions, both phenomena of ongoing star
formation. With the low density of faint radio sources, it is unlikely to
have a chance association within a few arc second \citep[see e.g.][ and
references therein]{Ivison2007}. Unfortunately, no sensitive radio
observation is available to date on the observed field.  Nevertheless,
Mid-IR imaging with e.g. Spitzer has also been used to identify SMGs
\citep{Pope2006,Ivison2007}, although the position uncertainties are larger
than for the radio observation and the link between Mid-IR and Far-IR
emission is not as well defined as the one between Far-IR and radio
\citep[see e.g.][]{Bavouzet2007} . We used data taken with IRAC (at 3.6,
4.5, 5.8 and 8.0~\micron) (Colbert et al. in prep.) and MIPS (at 24~\micron)
\citep{Colbert2006} to identify the sources detected at 870~\micron. From
the 3.6~\micron\ catalog we derived the photometry in all IRAC bands and
matched it to the 24~\micron\ catalog within a search radius of 2\arcsec\ to
produce a global search catalog. Stars were removed using a simple color
criterion of $S_{24\,\micron}/S_{3.6\,\micron} < 0.1$
\citep{Rodighiero2006}.

Following \citet{Condon1997} and \citet{Ivison2007}, the positional
uncertainties of a three free parameter Gaussian fit is $\Delta \alpha =
\Delta \delta = \theta\ (S/N)^{-1} / {2 \sqrt{\ln{2}}}$, where $\theta$ is
the beam HPBW and $(S/N)$ is the flux signal to noise ratio.  Adopting a
conservative lower value of $(S/N) = 3$, the rms position uncertainties are
on the order of 4\arcsec\ in both axes. Taking into account a possible shift
between Spitzer and \laboca\ absolute astrometric frames on the order of
$5\arcsec$, we have adopted a final search radius of $r=8\arcsec$.

\begin{figure*}
  \centering
  \includegraphics[width=17cm]{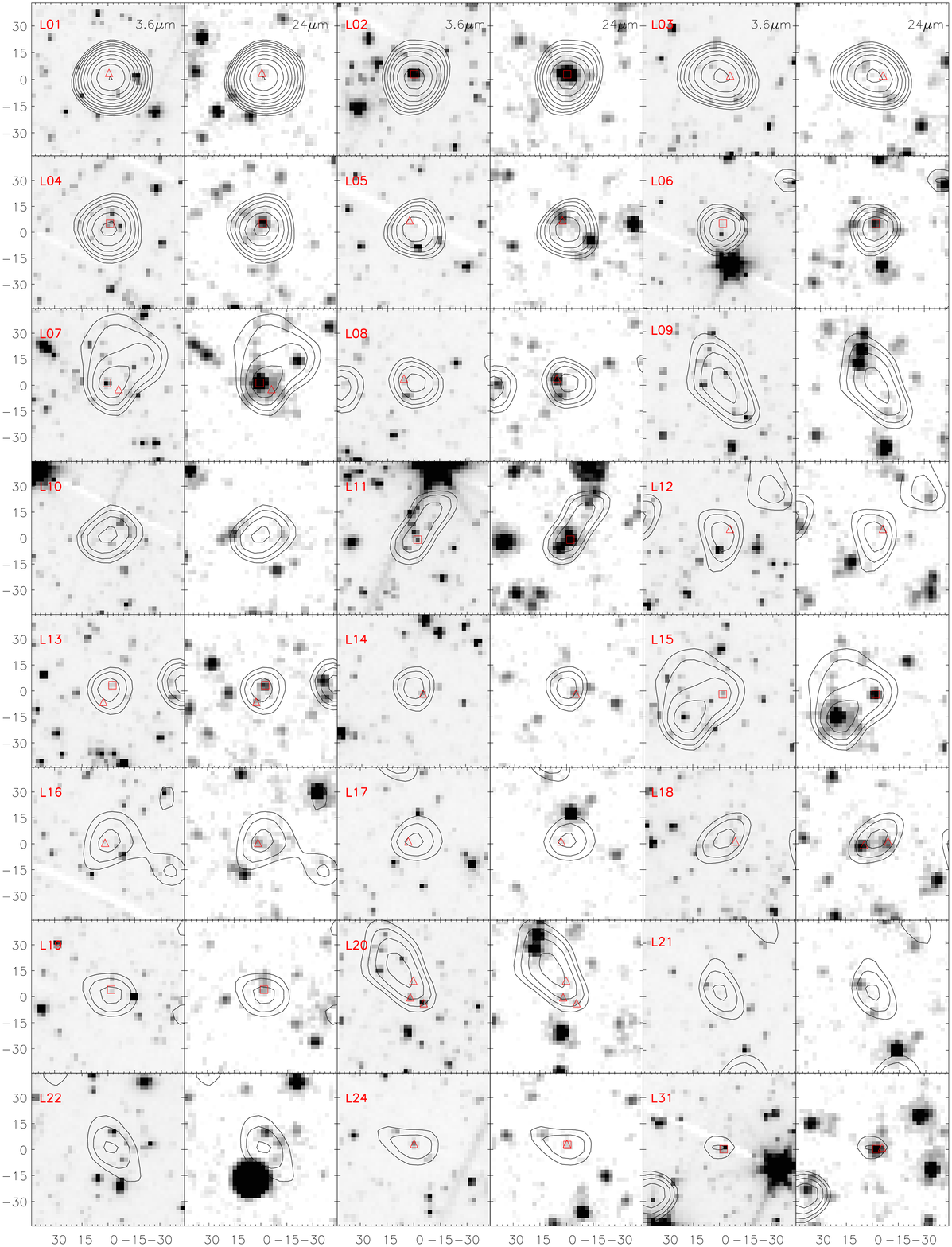}
  \caption{Spitzer images of the sources detected at 870~\micron with
    $\Delta F/F>1.2$ plus L24 and L31. From left to right, IRAC 3.6~\micron\
    and MIPS 24~\micron. Contours show the 870~\micron\ detection threshold
    $F/\Delta F$ at 0.8, 1.0, 1.2, ..., 5.0 in exponential progression. The
    axes denote the offsets in arc seconds from the 870~\micron\ source
    position. Sources are labeled in the top left corner (see text and
    Tables~\ref{tab:sources} and \ref{tab:id}). Squares show sources
    identified in Tab.~\ref{tab:id} with $P<0.05$ whereas triangles show
    identifications with $P>0.05$. }
  \label{fig:stamps}
\end{figure*}

For each 870~\micron\ source, we searched for all counterparts in the full
Mid-IR catalog and we computed the corrected Poisson probability of a chance
association $P$ using the method described in \citet{Downes1986} with the
number counts at 24~\micron\ and 3.6~\micron\ of
\citet{Papovich2004,Lagache2004} and \citet{Fazio2004}, the latter with IRAC
on Spitzer. The position and flux densities of all 24 and 3.6~\micron\
counterparts within 8\arcsec of the 870~\micron\ \laboca\ sources are
presented in Table~\ref{tab:id}, where the most reliable counterparts, with
$P<0.05$, are listed in bold face. From the 22+12 detected SMGs, ten show a
robust counterpart at 24~\micron, 15 others have a 24~\micron\ counterpart
with a probability $P\lesssim 0.2$ of spurious association, and finally ten
sources do not show a counterpart within 8\arcsec.  Practically all these
24~\micron\ sources but one, have a secure IRAC associate (within $\lesssim
1\arcsec$) detected at least at 3.6~\micron\ (Table~\ref{tab:id}).
Figure~\ref{fig:stamps} presents $45\arcsec\times45\arcsec$ postage stamp
images of the 3.6 and 24~\micron\ emission centered on the 22 securely
detected SMGs with $F/\Delta F > 1.2$. The lack of radio data over that
field leads to only half of the sources having a secure counterpart, whereas
it is of order of two-thirds in other submillimeter studies
\citep{Pope2006,Ivison2007} in which source positions are validated by radio
identifications. If one takes into account less secure associations
(Tab.~\ref{tab:id}), the total proportion of 24~\micron\ associations become
comparable.

From the four 50~kpc-Ly$\alpha$ blobs present in the observed field
\citep{Palunas2004}, of which three are detected at 24~\micron\
\citep{Colbert2006}, only one, B7, is detected at 870~\micron. None of the
other Ly$\alpha$ emitters present in the 870~\micron\ image is securely
detected, placing an upper limit on their flux densities of
$S_\mathrm{870\,\micron} \lesssim 5-7\, \mathrm{mJy}$ ($3\sigma$). However,
there are hints of emission near the Ly$\alpha$ blobs B4 and B6.

\begin{figure*}
  \centering
  \includegraphics[width=17cm]{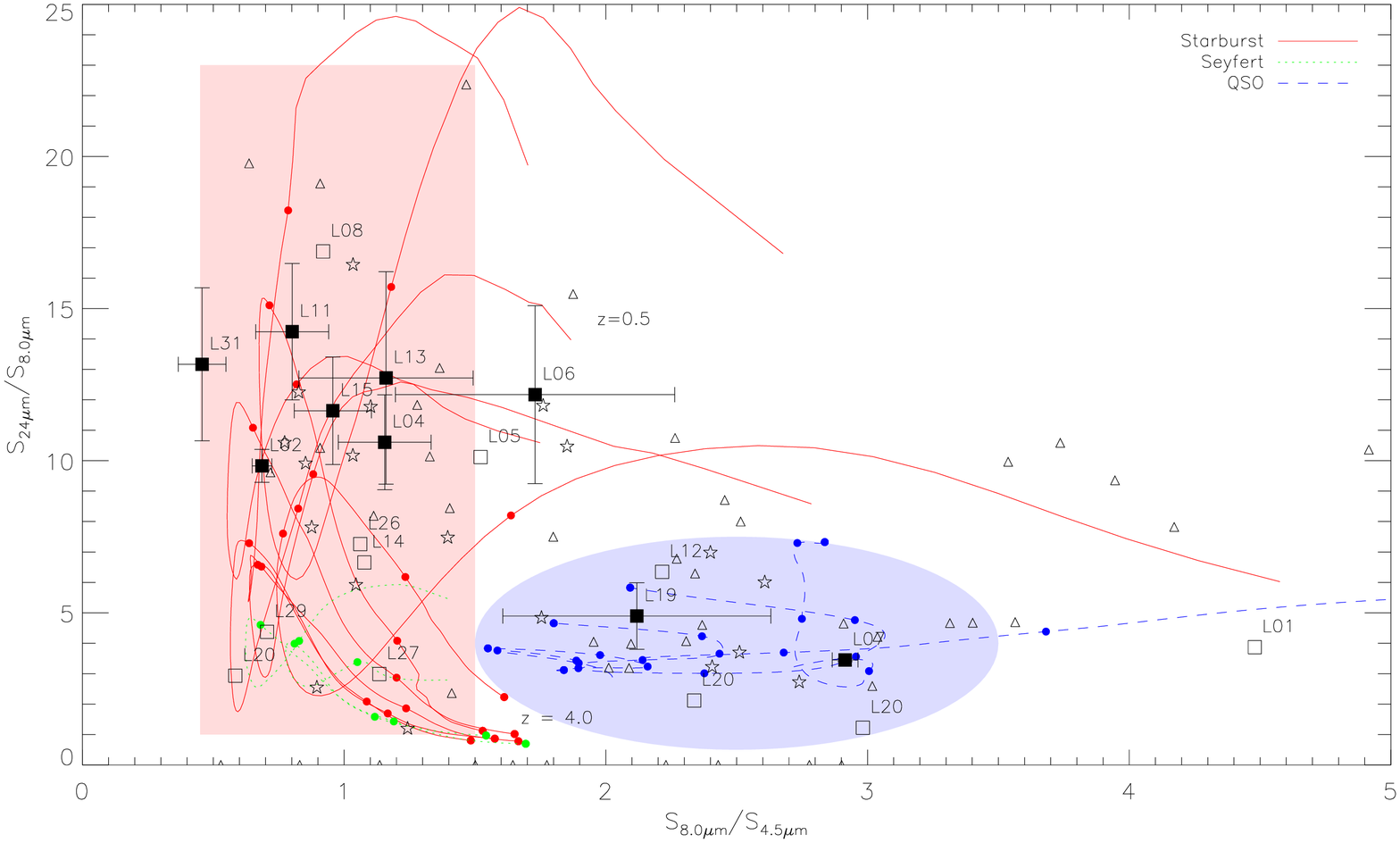}
  \caption{$S_\mathrm{24\, \micron}/S_\mathrm{8\, \micron}$ versus
    $S_\mathrm{8\, \micron}/S_\mathrm{4.5\, \micron}$ color-color diagram
    for the SMGs of Tab.~\ref{tab:id} with $P<0.05$ as full square, and open
    square for those with $P>0.05$. The lines are color-color tracks with
    redshift starting from 0.5 to 4, dots denote redshift 1 to 4, from top
    to bottom for the starburst-like SEDs (plain lines), Seyfert Galaxies
    (dotted lines) and QSOs (dashed lines). The rectangle defines the region
    where we expect most of the starbursts, whereas the ellipse shows the
    region of QSOs. Stars are the submillimeter galaxies of \citet{Pope2006}
    with secure Mid-IR counterparts, triangles are the radio galaxies of
    \citet{Seymour2007}.}
  \label{fig:colorcolor}
\end{figure*}

Mid-IR colors can be used to distinguish between starburst and AGN emission
\citep{Ivison2004,Pope2006,Ivison2007}. We compared the
24~\micron/8.0~\micron\ and 8.0~\micron/4.5~\micron\ color of the 9 sources
with robust and complete Mid-IR counterpart, with the SWIRE template SED
library of typical galaxies described in \citet{Polletta2007}. We used 14
template SEDs consisting of 6 starbursts corresponding to the SED of
Arp~220, M~82, NGC~6090, NGC~6240, IRAS~22491-1808, and IRAS~20551-4250, and
6 QSO-like SED, 3 type-1 and 2 type-2 and Mrk~231. We also included 2
moderately luminous AGN representing Seyfert 1.8 and Seyfert 2 galaxies
\citep[see][ for a detailed description]{Polletta2007}. These templates were
redshifted from $z=0.5$ to 4 and flux densities where computed using the
IRAC and MIPS filters. The results are presented in
Fig.~\ref{fig:colorcolor}, for all the sources in Tab.~\ref{tab:id},
together with sources with secure counterparts of \citet{Pope2006} and
\citet{Seymour2007}. From this diagram, it appears that 6/9 sources (L02,
L04, L11, L13, L15 and L33) have colors similar to starburst galaxies at
redshift between 1 and 3, whereas 2/9 (L07 and L11) are more likely to be
quasars. Finally, one, L06 lies close to the starburst region (see
Fig.\ref{fig:colorcolor}).

\subsection{Notes on Individual Objects}
\label{sec:notes}

\emph{L01} -- This source is among the strongest submillimeter sources and
is not securely associated with any Mid-IR source. No radio emission at
843~MHz is detected at the source position in the Sydney University Molonglo
Sky Survey (SUMSS) \citep{Mauch2003} placing an upper limit on the radio
flux density of 10~mJy. The submm-to-radio spectral index is then
$\alpha^\mathrm{345 GHz}_\mathrm{843 MHz} > 0.13$ which places this source
at $z> 0.7$ following the submm-to-radio spectral index relation of
\citet{Carilli2000} with the spectrum of M82 as template. The possible
Spitzer counterpart reported in Tab.~\ref{tab:id} has a IR SED typical of an
AGN. However, the very large $S_{870\,\micron}/S_{24\,\micron}$ ratio is
incompatible with an AGN, but rather needs a strong starburst SED such as
that of Arp~220. It is therefore possible that the main Spitzer counterpart
is undetected. With a secure completeness limit of $S_{24\,\micron} <
80\,\mathrm{\mu Jy}$, using the limit for the
$S_{870\,\micron}/S_{24\,\micron}$ ratio $(> 270)$, we can estimate a lower
limit on the source's redshift at $z \gtrsim 2.8$ (although $z = 2.38$
remains possible with a highly obscured SED). It is even possible that this
source is extended or associated with multiple objects within the large
\laboca\ beam, like other very strong submillimeter sources such as
SMM~J123711.7+622212 (GN20) \citep{Pope2006, Iono2006} or LAB1-SA~22
\citep{Matsuda2007}. Using a modified blackbody with a dust temperature of
40~K and a spectral index of 1.5, this corresponds to an infrared luminosity
of $\mathrm{L_{IR}} = \mathrm{L_{5-1000\,\micron}} \approx 2\, 10^{13}\,
\mathrm{L_\odot}$.

\emph{L02} -- This source has a very strong 24~\micron\ flux and Mid-IR
colors well within the starburst region in Fig.~\ref{fig:colorcolor}. There
is a $S_\mathrm{843 MHz} = 18.6\pm 1.3\, \mathrm{mJy}$ radio source at
9.6\arcsec\ displaced from our nominal position in the SUMSS Catalog. Due to
the large beam of the radio observation, it is possible that these two
sources are associated, making L02 a radio loud source. Its SED would then
be similar to e.g.\ 4C~41.17 \citep{Seymour2007, Archibald2001} but at much
lower redshift $z \sim 1.0-1.3$, compatible with the
$S_{870\,\micron}/S_{24\,\micron}$ ratio of 9. However given this low ratio,
this source could also be at higher redshift, up to 2.5.

\emph{L04} -- From the IRAC color of this source, we infer a starburst
template (Fig.~\ref{fig:colorcolor}) with a redshifted stellar maximum
emission in the 5.8~\micron\ band and an infrared redshift of $z_\mathrm{IR}
= 2.5\pm0.3$ inferred from a combination of the four IRAC fluxes
\citep{Pope2006} . Using Arp~220 or IRAS~20551-4250 as best matched
templates, we derive an infrared luminosity of $\mathrm{L_{IR}} \approx 1.4
\, 10^{13}\, \mathrm{L_\odot}$.

\emph{L06} -- This source is associated with the Ly$\alpha$ Blob B7 at
$z=2.38$ of \citet{Palunas2004}. This is one of the sources with the largest
excess flux in the narrow band Ly$\alpha$ filter compared to the B filter.
It has a very good 24~\micron\ association (Colbert et al.\ 2006).  There
appear to be additional MIPS sources at short distance of 15~\arcsec. The
SED of L06 is similar to IRAS~19254-7245 South, a Seyfert 2 and
Starbust/ULIRG composite leading to $\mathrm{L_{IR}} \approx 5 \, 10^{12}\
\mathrm{L_\odot}$ placing this source in the ULIRG regime. With a
Ly$\alpha$-IR relation of $\mathrm{L_{Ly\alpha}}/\mathrm{L_{bol}} = 0.16\%$,
this source lies on the trend found by \citet{Geach2005}.

\emph{L07} -- This source is associated with the QSO \object{[FPT2004]
  J214251.50-443043.2} detected by \citet{Francis2004} at $z=1.795$ with a B
magnitude of $M_B=20.26$. The SED of L07 is well described by a typical type
1 QSO SED with high IR counterpart, with an inferred luminosity of
$\mathrm{L_{IR}} = 8\, 10^{12}\, \mathrm{L_\odot}$.

\emph{L11} -- As for L04, the SED rather indicates a starburst
(Fig.~\ref{fig:colorcolor}) with a stellar bump in the 5.8\micron\ band. The
infrared redshift of this source is $z_\mathrm{IR} = 2.9\pm0.3$. Using this
redshift and the SED of IRAS~20551-4250, we derive an infrared luminosity of
$\mathrm{L_{IR}} \approx 2\, 10^{13}\, \mathrm{L_\odot}$.

\emph{L13} -- Having a similar SED to L04, this source has a infrared
redshift of $z_\mathrm{IR} = 2.7\pm0.3$, and an inferred infrared
luminosity of $\mathrm{L_{IR}} \approx 8\, 10^{12}\, \mathrm{L_\odot}$.

\emph{L15} -- This source has a typical starburst SED like IRAS~20551-4250
with an infrared redshift of $z_\mathrm{IR} = 2.8\pm0.3$ and luminosity of
$\mathrm{L_{IR}} \approx 1\, 10^{13}\, \mathrm{L_\odot}$. If this redshift
estimate is right, the proximity on the sky of L07 and L15
(Figs.~\ref{fig:s2n} and \ref{fig:stamps}) is fortuitous.

\emph{L19} -- If the association of Tab.~\ref{tab:id} is correct, the mid-IR
SED and colors are typical of an AGN, with an uncertain redshift, probably
$z \sim$ 2--3.


Nine sources have at least one counterpart within the search radius of
8\arcsec\ and within the completeness limit of
$S_{24\,\micron}\sim80\,\mathrm{\mu Jy}$, but the associations are less
secure with probabilities of spurious association in the range 0.05--0.20.
For three of them, L03, L05 L14, the $S_{870\,\micron}/S_{24\,\micron}$
ratio, greater than 135, indicates most likely starburst galaxies and places
them at $z \gtrsim 2.5$, but they could be at a lower redshift if their SEDs
where strongly obscured.  For the six other galaxy candidates, L08, L12,
L16, L17, L18, L20, the $S_{870\,\micron}/S_{24\,\micron}$ ratio, ranging
from $>35$ to $>95$, places them at $z>2.2$, or around $z\sim1.4$ if they
have obscured SEDs similar to Arp~220. In the case of L12, the tentative
counterpart presents typical colors of an AGN, which would place it at very
high redshift ($z>5$).


Two sources, L09 and L10, have no counterpart candidates within the
completeness limit of the 24~\micron\ observations. With respective lower
limits for the $S_{870\,\micron}/S_{24\,\micron}$ ratio of 140 and 120,
these sources are likely to be starbursts at $z>2.5$ or at lower redshift in
case of a strongly obscured starburst SED. L09 is close to the position of
the Ly$\alpha$ Blob B6, but the blending with L20 makes the association
unlikely. Last, L21 has also no counterpart candidates but with a lower
$S_{870\,\micron}$ flux density, this source's redshift is more likely to be
around 1.5.


The quality of the submillimeter fluxes of the ten other sources, L22-32 of
Tab.~\ref{tab:sources} \& \ref{tab:id}, is lower, as well as most of their
possible Spitzer associations.  Exceptions of L24 which has a high
probability 24~\micron\ counterpart (see Fig.~\ref{fig:stamps}), with
$S_{870\,\micron}/S_{24\,\micron} \sim 55 $, indicating a starburst at
$z\sim 1.5$, L25 which lies close to the Ly$\alpha$ Blob B4, but the higher
noise on the edge of the map prevents any firm conclusion and L31 which
seems to lie on the edge of the starburst region on the color-color diagram
presented Fig.~\ref{fig:colorcolor} and has $z_\mathrm{IR} = 2.7\pm0.3$.

\section{Discussion \& Conclusions}
\label{sec:final}

By detecting high-z powerful ULIRGs ($L_{FIR} \sim 5-20 \, 10^{12}\,
\mathrm{L_\odot}$), our results may cast some light on the association of
such SMGs not only with the Ly$\alpha$ blobs themselves, but also with the
$z=2.38$ large scale structure that they trace. The meager submillimeter
detection rate of only one Ly$\alpha$ blob confirms the findings of
\citet{Geach2005} that maximum starbursts with $L_{FIR} \gtrsim 10^{13}\,
\mathrm{L_\odot}$ are present in only a minority of even 50~kpc-Ly$\alpha$
blobs. The detection rate of one out of four 50~kpc-Ly$\alpha$ blobs and
none in the three other blobs may appear significantly lower than in those
of SA~22 by \citet{Geach2005} who detected three out of ten
50~kpc-Ly$\alpha$ blobs, and two more among 15 additional moderately
extended ($\sim 35\,\mathrm{kpc}$) blobs. But one should also take into
account that, except the detected blob L06/B7, all other blobs are located
in regions of the map with relatively high noise, $\sim
2.0-2.7\,\mathrm{mJy}$, i.e.  larger than the rms of the detections of B7 or
of \citet{Geach2005} by factors~$\gtrsim 1.5$. Moreover there is a $2\sigma$
870~\micron\ detection ($4.9\pm2.0\,\mathrm{mJy}$) at the 24~\micron\
position of the B6 blob of \citet{Colbert2006}. There are two 870~\micron\
sources, L09 and L20, within $\sim30\arcsec$ of B6. The average values of
the 870~\micron\ flux densities for undetected blobs --
$2.3\pm0.8\,\mathrm{mJy}$ for the 50~kpc-blobs B1, B5 \& B6, and
$1.2\pm0.6\,\mathrm{mJy}$ for all, B1, B2, B4, B5, B6 \& B8 -- are also
compatible with the value of \citet{Geach2005}, $1.2\pm0.4\,\mathrm{mJy}$
for all their blobs. Nevertheless, the non detection at 870~\micron\ of the
blobs B4 and B6 excludes the very large far-IR luminosities, $4-5 \,
10^{13}\, \mathrm{L_\odot}$, inferred by \citet{Colbert2006}, placing upper
limits rather $\lesssim 10^{13}\, \mathrm{L_\odot}$.

More important, we have detected at least twenty additional sources at
870~\micron, with indications that several of them could be associated with
the $z=2.38$ LSS traced by the Ly$\alpha$ blobs. As discussed in
Sect.~\ref{sec:notes}, out of 21 most secure detections outside of L06/B7, 7
sources, L03, L04 L05, L09, L10, L13 \& L15, with starburst-like SED have
probable redshifts in the range 2.0--2.8 compatible with the redshift of the
galaxies protocluster J2143-4423 at $z=2.38$; moreover seven additional
sources, L01, L08, L12, L16, L17, L18 \& L20 have more uncertain redshifts,
but still not incompatible with $z=2.38$. Only L02 and L07 have clear
indication of lower redshifts. As quoted in Sect.~\ref{sec:counts}, five
detected sources, are within a circle of diameter 5\arcmin\ centred on B7,
yielding an apparent strong over-density (Fig.~\ref{fig:counts}), which is
another argument for their possible membership of a protocluster at
$z=2.38$. However, spectroscopic determination of the redshifts will be
mandatory to confirm the existence of such a compact protocluster and the
membership of these SMGs.

The possible redshifts of our submillimeter detections, with most of the
sources probably at $z \gtrsim 2$ are in agreement with the general
distribution of the redshifts of SMGs \citep{Chapman2005}.  However,
regarding the nature of the sources, we note a trend for a larger proportion
($\sim 30\%$) of AGN-starburst composite mid-IR SEDs than is usually found
\citep[e.g.][]{Valiante2007,Pope2007}: L06/B7 itself has clearly such a
profile, as well as L07 and L19 and possibly L12 and L20; there is a
possible match of L02 with a radio galaxy although its $3.6-8.0\,\micron$
SED looks starburst dominated at $z \sim 1.0-1.3$. As discussed, the case of
one of the strongest submillimeter sources known to date, L01
($21\,\mathrm{mJy}$) is not completely clear since the mid-IR SED of a
possible Spitzer counterpart looks AGN dominated, while the
870~\micron/24~\micron\ flux ratio suggests a starburst-dominated profile.

This first published deep map illustrates and confirms the capabilities of
the \laboca\ camera on APEX. As expected, it is currently the most efficient
equipment for wide and deep submillimeter mapping. The multiple checks that
we have performed about the overall quality of the map, of the achieved
signal-to-noise ratio and of the source extraction demonstrate the
efficiency of the observing mode and of the data processing which were used
and optimized for this work.

The first priority for further work aimed at confirming the nature of these
SMGs and their possible relationship with the Ly$\alpha$ blobs and their
underlying structure, is the determination of their redshifts. Both optical
and near-IR determinations should be possible if the redshifts are in the
range of $z=2.38$, and they should be especially easy for AGN. More complete
optical and near-IR data could be useful for this purpose and for
considering the possibility of adaptive optics studies to trace the source's
inner structures. Deep radio data (with ATCA, VLA and/or GMRT) would be
important to trace the star formation rates and if possible the extension of
starbursts, as well as the radio spectral index for trying to disentangle
AGN and starburst contributions. Stacking 870~\micron\ intensities of radio
and Spitzer 24~\micron\ and IRAC sources will extend the information about
source counts and the average star formation rate to weaker submillimeter
sources. In parallel, a deeper or wider \laboca\ map would be useful to
increase the number of individual detections, especially of Ly$\alpha$
blobs, and the stacking accuracy. High resolution (sub)millimeter imaging
would be essential to check the spatial structure, especially of the
strongest source L01, similarly to the observation of LAB01-SA22 by
\citet{Matsuda2007}. Waiting for ALMA, the J2143-4423 region is just at the
limit of observability with SMA.

\begin{acknowledgements}
  We thank the referee for comments, which improved the manuscript.  We
  thank F.~Schuller for help during the development of the BoA software. AB
  thanks M.~Douspis and M.~Langer for their useful interactions during the
  writing of this paper. This work is based on observations made with the
  APEX Telescope and the \emph{Spitzer Space Telescope}. APEX is a
  collaboration between the Max-Planck-Institut f\"ur Radioastronomie, the
  European Southern Observatory, and the Onsala Space Observatory. The
  \emph{Spitzer Space Telescope} is operated by the Jet Propulsion
  Laboratory, California Institute of Technology under NASA contract
  1407. We thank the APEX staff and astronomers for their support during the
  observations and the Bolometer Development Group at the MPIfR for
  providing the \laboca\ bolometer array.  Last, but not least, AB thanks
  the staff of Adobe and Cafe Export in San Pedro de Atacama.
\end{acknowledgements}

\bibliographystyle{aa} 
\bibliography{9500}   

\begin{thebibliography}{40}
\expandafter\ifx\csname natexlab\endcsname\relax\def\natexlab#1{#1}\fi

\bibitem[{{Archibald} {et~al.}(2001){Archibald}, {Dunlop}, {Hughes},
  {Rawlings}, {Eales}, \& {Ivison}}]{Archibald2001}
{Archibald}, E.~N., {Dunlop}, J.~S., {Hughes}, D.~H., {et~al.} 2001, \mnras,
  323, 417

\bibitem[{{Bavouzet} {et~al.}(2007){Bavouzet}, {Dole}, {Le Floc'h}, {Caputi},
  {Lagache}, \& {Kochanek}}]{Bavouzet2007}
{Bavouzet}, N., {Dole}, H., {Le Floc'h}, E., {et~al.} 2007, ArXiv e-prints, 712

\bibitem[{{Carilli} \& {Yun}(2000)}]{Carilli2000}
{Carilli}, C.~L. \& {Yun}, M.~S. 2000, \apj, 539, 1024

\bibitem[{{Chapman} {et~al.}(2005){Chapman}, {Blain}, {Smail}, \&
  {Ivison}}]{Chapman2005}
{Chapman}, S.~C., {Blain}, A.~W., {Smail}, I., \& {Ivison}, R.~J. 2005, \apj,
  622, 772

\bibitem[{{Colbert} {et~al.}(2006){Colbert}, {Teplitz}, {Francis}, {Palunas},
  {Williger}, \& {Woodgate}}]{Colbert2006}
{Colbert}, J.~W., {Teplitz}, H., {Francis}, P., {et~al.} 2006, \apjl, 637, L89

\bibitem[{{Condon}(1997)}]{Condon1997}
{Condon}, J.~J. 1997, \pasp, 109, 166

\bibitem[{{Coppin} {et~al.}(2006){Coppin}, {Chapin}, {Mortier}, {Scott},
  {Borys}, {Dunlop}, {Halpern}, {Hughes}, {Pope}, {Scott}, {Serjeant}, {Wagg},
  {Alexander}, {Almaini}, {Aretxaga}, {Babbedge}, {Best}, {Blain}, {Chapman},
  {Clements}, {Crawford}, {Dunne}, {Eales}, {Edge}, {Farrah}, {Gazta{\~n}aga},
  {Gear}, {Granato}, {Greve}, {Fox}, {Ivison}, {Jarvis}, {Jenness}, {Lacey},
  {Lepage}, {Mann}, {Marsden}, {Martinez-Sansigre}, {Oliver}, {Page},
  {Peacock}, {Pearson}, {Percival}, {Priddey}, {Rawlings}, {Rowan-Robinson},
  {Savage}, {Seigar}, {Sekiguchi}, {Silva}, {Simpson}, {Smail}, {Stevens},
  {Takagi}, {Vaccari}, {van Kampen}, \& {Willott}}]{Coppin2006}
{Coppin}, K., {Chapin}, E.~L., {Mortier}, A.~M.~J., {et~al.} 2006, \mnras, 372,
  1621

\bibitem[{{Dobbs} {et~al.}(2006){Dobbs}, {Halverson}, {Ade}, {Basu},
  {\textbf{Beelen}}, {Bertoldi}, {Cohalan}, {Cho}, {G{\"u}sten}, {Holzapfel},
  {Kermish}, {Kneissl}, {Kov{\'a}cs}, {Kreysa}, {Lanting}, {Lee}, {Lueker},
  {Mehl}, {Menten}, {Muders}, {Nord}, {Plagge}, {Richards}, {Schilke},
  {Schwan}, {Spieler}, {Weiss}, \& {White}}]{Dobbs2006}
{Dobbs}, M., {Halverson}, N.~W., {Ade}, P.~A.~R., {et~al.} 2006, New Astronomy
  Review, 50, 960

\bibitem[{{Downes} {et~al.}(1986){Downes}, {Peacock}, {Savage}, \&
  {Carrie}}]{Downes1986}
{Downes}, A.~J.~B., {Peacock}, J.~A., {Savage}, A., \& {Carrie}, D.~R. 1986,
  \mnras, 218, 31

\bibitem[{{Fazio} {et~al.}(2004){Fazio}, {Ashby}, {Barmby}, {Hora}, {Huang},
  {Pahre}, {Wang}, {Willner}, {Arendt}, {Moseley}, {Brodwin}, {Eisenhardt},
  {Stern}, {Tollestrup}, \& {Wright}}]{Fazio2004}
{Fazio}, G.~G., {Ashby}, M.~L.~N., {Barmby}, P., {et~al.} 2004, \apjs, 154, 39

\bibitem[{{Francis} {et~al.}(2004){Francis}, {Palunas}, {Teplitz}, {Williger},
  \& {Woodgate}}]{Francis2004}
{Francis}, P.~J., {Palunas}, P., {Teplitz}, H.~I., {Williger}, G.~M., \&
  {Woodgate}, B.~E. 2004, \apj, 614, 75

\bibitem[{{Francis} {et~al.}(1997){Francis}, {Woodgate}, \&
  {Danks}}]{Francis1997}
{Francis}, P.~J., {Woodgate}, B.~E., \& {Danks}, A.~C. 1997, \apjl, 482, L25+

\bibitem[{{Francis} {et~al.}(1996){Francis}, {Woodgate}, {Warren}, {Moller},
  {Mazzolini}, {Bunker}, {Lowenthal}, {Williams}, {Minezaki}, {Kobayashi}, \&
  {Yoshii}}]{Francis1996}
{Francis}, P.~J., {Woodgate}, B.~E., {Warren}, S.~J., {et~al.} 1996, \apj, 457,
  490

\bibitem[{{Geach} {et~al.}(2005){Geach}, {Matsuda}, {Smail}, {Chapman},
  {Yamada}, {Ivison}, {Hayashino}, {Ohta}, {Shioya}, \&
  {Taniguchi}}]{Geach2005}
{Geach}, J.~E., {Matsuda}, Y., {Smail}, I., {et~al.} 2005, \mnras, 363, 1398

\bibitem[{{G{\"u}sten} {et~al.}(2006){G{\"u}sten}, {Nyman}, {Schilke},
  {Menten}, {Cesarsky}, \& {Booth}}]{Gusten2006}
{G{\"u}sten}, R., {Nyman}, L.~{\AA}., {Schilke}, P., {et~al.} 2006, \aap, 454,
  L13

\bibitem[{{H{\"o}gbom}(1974)}]{Hogbom1974}
{H{\"o}gbom}, J.~A. 1974, \aaps, 15, 417

\bibitem[{{Iono} {et~al.}(2006){Iono}, {Peck}, {Pope}, {Borys}, {Scott},
  {Wilner}, {Gurwell}, {Ho}, {Yun}, {Matsushita}, {Petitpas}, {Dunlop},
  {Elvis}, {Blain}, \& {Le Floc'h}}]{Iono2006}
{Iono}, D., {Peck}, A.~B., {Pope}, A., {et~al.} 2006, \apjl, 640, L1

\bibitem[{{Ivison} {et~al.}(2007){Ivison}, {Greve}, {Dunlop}, {Peacock},
  {Egami}, {Smail}, {Ibar}, {van Kampen}, {Aretxaga}, {Babbedge}, {Biggs},
  {Blain}, {Chapman}, {Clements}, {Coppin}, {Farrah}, {Halpern}, {Hughes},
  {Jarvis}, {Jenness}, {Jones}, {Mortier}, {Oliver}, {Papovich},
  {P{\'e}rez-Gonz{\'a}lez}, {Pope}, {Rawlings}, {Rieke}, {Rowan-Robinson},
  {Savage}, {Scott}, {Seigar}, {Serjeant}, {Simpson}, {Stevens}, {Vaccari},
  {Wagg}, \& {Willott}}]{Ivison2007}
{Ivison}, R.~J., {Greve}, T.~R., {Dunlop}, J.~S., {et~al.} 2007, \mnras, 380,
  199

\bibitem[{{Ivison} {et~al.}(2004){Ivison}, {Greve}, {Serjeant}, {Bertoldi},
  {Egami}, {Mortier}, {Alonso-Herrero}, {Barmby}, {Bei}, {Dole}, {Engelbracht},
  {Fazio}, {Frayer}, {Gordon}, {Hines}, {Huang}, {Le Floc'h}, {Misselt},
  {Miyazaki}, {Morrison}, {Papovich}, {P{\'e}rez-Gonz{\'a}lez}, {Rieke},
  {Rieke}, {Rigby}, {Rigopoulou}, {Smail}, {Wilson}, \& {Willner}}]{Ivison2004}
{Ivison}, R.~J., {Greve}, T.~R., {Serjeant}, S., {et~al.} 2004, \apjs, 154, 124

\bibitem[{{Kov\'acs}(2006)}]{Kovacs2006}
{Kov\'acs}, A. 2006, PhD thesis, Caltech

\bibitem[{{Kreysa} {et~al.}(1998){Kreysa}, {Gemuend}, {Gromke}, {Haslam},
  {Reichertz}, {Haller}, {Beeman}, {Hansen}, {Sievers}, \&
  {Zylka}}]{Kreysa1998}
{Kreysa}, E., {Gemuend}, H.-P., {Gromke}, J., {et~al.} 1998, in Presented at
  the Society of Photo-Optical Instrumentation Engineers (SPIE) Conference,
  Vol. 3357, Proc. SPIE Vol. 3357, p. 319-325, Advanced Technology MMW, Radio,
  and Terahertz Telescopes, Thomas G. Phillips; Ed., ed. T.~G. {Phillips},
  319--325

\bibitem[{{Lagache} {et~al.}(2004){Lagache}, {Dole}, {Puget},
  {P{\'e}rez-Gonz{\'a}lez}, {Le Floc'h}, {Rieke}, {Papovich}, {Egami},
  {Alonso-Herrero}, {Engelbracht}, {Gordon}, {Misselt}, \&
  {Morrison}}]{Lagache2004}
{Lagache}, G., {Dole}, H., {Puget}, J.-L., {et~al.} 2004, \apjs, 154, 112

\bibitem[{{Matsuda} {et~al.}(2007){Matsuda}, {Iono}, {Ohta}, {Yamada},
  {Kawabe}, {Hayashino}, {Peck}, \& {Petitpas}}]{Matsuda2007}
{Matsuda}, Y., {Iono}, D., {Ohta}, K., {et~al.} 2007, \apj, 667, 667

\bibitem[{{Matsuda} {et~al.}(2004){Matsuda}, {Yamada}, {Hayashino}, {Tamura},
  {Yamauchi}, {Ajiki}, {Fujita}, {Murayama}, {Nagao}, {Ohta}, {Okamura},
  {Ouchi}, {Shimasaku}, {Shioya}, \& {Taniguchi}}]{Matsuda2004}
{Matsuda}, Y., {Yamada}, T., {Hayashino}, T., {et~al.} 2004, \aj, 128, 569

\bibitem[{{Mauch} {et~al.}(2003){Mauch}, {Murphy}, {Buttery}, {Curran},
  {Hunstead}, {Piestrzynski}, {Robertson}, \& {Sadler}}]{Mauch2003}
{Mauch}, T., {Murphy}, T., {Buttery}, H.~J., {et~al.} 2003, \mnras, 342, 1117

\bibitem[{{Miley} \& {De Breuck}(2008)}]{Miley2008}
{Miley}, G. \& {De Breuck}, C. 2008, \aapr, 15, 67

\bibitem[{{Nilsson} {et~al.}(2006){Nilsson}, {Fynbo}, {M{\o}ller},
  {Sommer-Larsen}, \& {Ledoux}}]{Nilsson2006}
{Nilsson}, K.~K., {Fynbo}, J.~P.~U., {M{\o}ller}, P., {Sommer-Larsen}, J., \&
  {Ledoux}, C. 2006, \aap, 452, L23

\bibitem[{{Palunas} {et~al.}(2004){Palunas}, {Teplitz}, {Francis}, {Williger},
  \& {Woodgate}}]{Palunas2004}
{Palunas}, P., {Teplitz}, H.~I., {Francis}, P.~J., {Williger}, G.~M., \&
  {Woodgate}, B.~E. 2004, \apj, 602, 545

\bibitem[{{Papovich} {et~al.}(2004){Papovich}, {Dole}, {Egami}, {Le Floc'h},
  {P{\'e}rez-Gonz{\'a}lez}, {Alonso-Herrero}, {Bai}, {Beichman}, {Blaylock},
  {Engelbracht}, {Gordon}, {Hines}, {Misselt}, {Morrison}, {Mould},
  {Muzerolle}, {Neugebauer}, {Richards}, {Rieke}, {Rieke}, {Rigby}, {Su}, \&
  {Young}}]{Papovich2004}
{Papovich}, C., {Dole}, H., {Egami}, E., {et~al.} 2004, \apjs, 154, 70

\bibitem[{{Polletta} {et~al.}(2007){Polletta}, {Tajer}, {Maraschi},
  {Trinchieri}, {Lonsdale}, {Chiappetti}, {Andreon}, {Pierre}, {Le Fevre},
  {Zamorani}, {Maccagni}, {Garcet}, {Surdej}, {Franceschini}, {Alloin},
  {Shupe}, {Surace}, {Fang}, {Rowan-Robinson}, {Smith}, \&
  {Tresse}}]{Polletta2007}
{Polletta}, M., {Tajer}, M., {Maraschi}, L., {et~al.} 2007, ArXiv Astrophysics
  e-prints

\bibitem[{{Pope} {et~al.}(2007){Pope}, {Chary}, {Alexander}, {Armus},
  {Dickinson}, {Elbaz}, {Frayer}, {Scott}, \& {Teplitz}}]{Pope2007}
{Pope}, A., {Chary}, R.-R., {Alexander}, D.~M., {et~al.} 2007, ArXiv e-prints,
  711

\bibitem[{{Pope} {et~al.}(2006){Pope}, {Scott}, {Dickinson}, {Chary},
  {Morrison}, {Borys}, {Sajina}, {Alexander}, {Daddi}, {Frayer}, {MacDonald},
  \& {Stern}}]{Pope2006}
{Pope}, A., {Scott}, D., {Dickinson}, M., {et~al.} 2006, \mnras, 370, 1185

\bibitem[{{Rodighiero} {et~al.}(2006){Rodighiero}, {Lari}, {Pozzi},
  {Gruppioni}, {Fadda}, {Franceschini}, {Lonsdale}, {Surace}, {Shupe}, \&
  {Fang}}]{Rodighiero2006}
{Rodighiero}, G., {Lari}, C., {Pozzi}, F., {et~al.} 2006, \mnras, 371, 1891

\bibitem[{{Serjeant} {et~al.}(2003){Serjeant}, {Dunlop}, {Mann},
  {Rowan-Robinson}, {Hughes}, {Efstathiou}, {Blain}, {Fox}, {Ivison},
  {Jenness}, {Lawrence}, {Longair}, {Oliver}, \& {Peacock}}]{Serjeant2003}
{Serjeant}, S., {Dunlop}, J.~S., {Mann}, R.~G., {et~al.} 2003, \mnras, 344, 887

\bibitem[{{Seymour} {et~al.}(2007){Seymour}, {Stern}, {De Breuck}, {Vernet},
  {Rettura}, {Dickinson}, {Dey}, {Eisenhardt}, {Fosbury}, {Lacy}, {McCarthy},
  {Miley}, {Rocca-Volmerange}, {R{\"o}ttgering}, {Stanford}, {Teplitz}, {van
  Breugel}, \& {Zirm}}]{Seymour2007}
{Seymour}, N., {Stern}, D., {De Breuck}, C., {et~al.} 2007, \apjs, 171, 353

\bibitem[{{Siringo} {et~al.}(2007){Siringo}, {Weiss}, {Kreysa}, {Schuller},
  {Kovacs}, {\textbf{Beelen}}, {Esch}, {Gem{\"u}nd}, {Jethava},
  {Lundershausen}, {Menten}, {G{\"u}sten}, {Bertoldi}, {De Breuck}, {Nyman},
  {Haller}, \& {Beeman}}]{Siringo2007}
{Siringo}, G., {Weiss}, A., {Kreysa}, E., {et~al.} 2007, The Messenger, 129, 2

\bibitem[{{Smith} \& {Jarvis}(2007)}]{Smith2007}
{Smith}, D.~J.~B. \& {Jarvis}, M.~J. 2007, \mnras, 378, L49

\bibitem[{{Steidel} {et~al.}(2000){Steidel}, {Adelberger}, {Shapley},
  {Pettini}, {Dickinson}, \& {Giavalisco}}]{Steidel2000}
{Steidel}, C.~C., {Adelberger}, K.~L., {Shapley}, A.~E., {et~al.} 2000, \apj,
  532, 170

\bibitem[{{Valiante} {et~al.}(2007){Valiante}, {Lutz}, {Sturm}, {Genzel},
  {Tacconi}, {Lehnert}, \& {Baker}}]{Valiante2007}
{Valiante}, E., {Lutz}, D., {Sturm}, E., {et~al.} 2007, \apj, 660, 1060

\bibitem[{{Yamada}(2007)}]{Yamada2007}
{Yamada}, T. 2007, http://sfig.pmo.ac.cn/xining/

\end{thebibliography}

\clearpage
\onecolumn
\begin{table}
\begin{minipage}[t]{17cm}
\caption{The 870~$\mathrm{\mu m}$ LABoCa source catalogue around the
  $z=2.38$ galaxy protocluster J2143-4423.}
\label{tab:sources}
\centering
\renewcommand{\footnoterule}{}  
\begin{tabular}{l l c c r r c r}
  \hline
  \hline
  Name & Nickname & $\alpha$ & $\delta$ &
  $S_{\nu}$ & 
  $F/\Delta F$\footnote{Gaussian matched-filtered detection threshold} &
  Notes\footnote{Identification of the source (see text)} \\
  (IAU) & & \multicolumn{2}{c}{(J2000)} & 
  [mJy]\footnote{Corrected from the flux boosting effect\label{note:boost}}
  \footnote{The absolute flux uncertainty is not included } & & \\
  \hline

\object{LABoCa J214239-442820} & L01 & 21 42 39.56 & -44 28 20.12 & $21.1\pm1.0$ & 5.22 &     \\
\object{LABoCa J214220-442454} & L02 & 21 42 20.98 & -44 24 54.22 & $13.7\pm1.2$ & 3.07 & SB  \\
\object{LABoCa J214248-442730} & L03 & 21 42 48.00 & -44 27 30.62 & $14.1\pm1.1$ & 2.90 &     \\
\object{LABoCa J214222-442813} & L04 & 21 42 22.08 & -44 28 13.16 & $12.0\pm1.1$ & 2.35 & SB  \\
\object{LABoCa J214231-442348} & L05 & 21 42 31.44 & -44 23 48.46 & $10.8\pm1.1$ & 2.16 &     \\
\object{LABoCa J214235-442711} & L06 & 21 42 35.03 & -44 27 11.99 & $ 8.4\pm1.0$ & 1.81 & B7  \\
\object{LABoCa J214251-443043} & L07 & 21 42 51.24 & -44 30 43.73 & $10.6\pm1.8$ & 1.75 & QSO \\
\object{LABoCa J214258-442501} & L08 & 21 42 58.69 & -44 25 01.70 & $ 7.6\pm1.1$ & 1.64 &     \\
\object{LABoCa J214241-443030} & L09 & 21 42 41.88 & -44 30 30.08 & $11.3\pm1.9$ & 1.60 &     \\
\object{LABoCa J214230-442813} & L10 & 21 42 30.54 & -44 28 13.34 & $ 8.9\pm1.0$ & 1.58 &     \\
\object{LABoCa J214244-442326} & L11 & 21 42 44.76 & -44 23 26.40 & $ 6.9\pm1.1$ & 1.41 & SB  \\
\object{LABoCa J214229-442215} & L12 & 21 42 29.30 & -44 22 15.30 & $ 6.3\pm1.2$ & 1.41 &     \\
\object{LABoCa J214302-442505} & L13 & 21 43 02.77 & -44 25 05.64 & $ 7.0\pm1.2$ & 1.40 & SB  \\
\object{LABoCa J214232-443127} & L14 & 21 42 32.30 & -44 31 27.24 & $14.0\pm2.5$ & 1.38 &     \\
\object{LABoCa J214249-443026} & L15 & 21 42 49.34 & -44 30 26.70 & $ 8.9\pm1.5$ & 1.37 & SB  \\
\object{LABoCa J214240-442517} & L16 & 21 42 40.68 & -44 25 17.04 & $ 4.3\pm0.8$ & 1.32 &     \\
\object{LABoCa J214209-442605} & L17 & 21 42 09.52 & -44 26 05.32 & $ 6.7\pm2.3$ & 1.31 &     \\
\object{LABoCa J214234-442208} & L18 & 21 42 34.25 & -44 22 08.35 & $ 3.0\pm1.0$ & 1.30 &     \\
\object{LABoCa J214216-442852} & L19 & 21 42 16.14 & -44 28 52.23 & $ 9.0\pm2.4$ & 1.29 & AGN \\
\object{LABoCa J214240-443045} & L20 & 21 42 40.86 & -44 30 45.31 & $ 8.1\pm1.6$ & 1.29 &     \\
\object{LABoCa J214210-442515} & L21 & 21 42 10.66 & -44 25 15.82 & $ 6.7\pm1.9$ & 1.26 &     \\
\object{LABoCa J214309-442825} & L22 & 21 43 09.39 & -44 28 25.24 & $11.1\pm2.9$ & 1.25 &     \\
\hline
\object{LABoCa J214244-442625} & L23 & 21 42 44.99 & -44 26 25.55 & $ 4.3\pm0.9$ & 1.19 &     \\
\object{LABoCa J214252-442457} & L24 & 21 42 52.75 & -44 24 57.77 & $ 5.2\pm0.9$ & 1.17 &     \\
\object{LABoCa J214232-442014} & L25 & 21 42 32.92 & -44 20 14.71 & $ 6.6\pm1.7$ & 1.15 &     \\
\object{LABoCa J214226-442149} & L26 & 21 42 26.56 & -44 21 49.64 & $ 5.0\pm1.3$ & 1.11 &     \\
\object{LABoCa J214311-442617} & L27 & 21 43 11.44 & -44 26 17.39 & $ 4.9\pm1.7$ & 1.08 &     \\
\object{LABoCa J214237-442534} & L28 & 21 42 37.38 & -44 25 34.77 & $ 2.9\pm0.8$ & 1.05 &     \\
\object{LABoCa J214304-442200} & L29 & 21 43 04.02 & -44 22 00.10 & $ 7.1\pm2.1$ & 1.04 &     \\
\object{LABoCa J214312-442736} & L30 & 21 43 12.32 & -44 27 36.62 & $ 4.8\pm2.4$ & 1.02 &     \\
\object{LABoCa J214231-442642} & L31 & 21 42 31.22 & -44 26 42.79 & $ 3.2\pm0.8$ & 1.02 & SB  \\
\object{LABoCa J214242-442210} & L32 & 21 42 43.00 & -44 22 10.13 & $ 4.1\pm1.0$ & 1.01 &     \\
\end{tabular}
\end{minipage}
\end{table}

\begin{table}
\begin{minipage}[t]{17cm}
  \caption{The Mid-IR properties of SMG around the $z=2.38$ galaxy protocluster
    J2143-4423.}
\label{tab:id}
\centering
\renewcommand{\footnoterule}{}  
\begin{tabular}{l c c r r r r r r r}
  \hline
  \hline
  Nickname & \multicolumn{5}{c}{MIPS 24~\micron} & 
  \multicolumn{4}{c}{IRAC} \\
  & R.A. & DEC  & $S_{\nu} \pm\sigma_{\nu}$  & Offset & P\footnote{P was
    computed with search radius of 8\arcsec. Reliable identifications ($P<0.05$) are listed in bold.}
  & $S_{3.6\,\micron}$ & $S_{4.5\,\micron}$ & $S_{5.8\,\micron}$ & $S_{8.0\,\micron}$ \\
  & \multicolumn{2}{c}{(J2000)} & [$\mathrm{\mu Jy}$] & \arcsec & 
  & [$\mathrm{\mu Jy}$] & [$\mathrm{\mu Jy}$] & [$\mathrm{\mu Jy}$] &
  [$\mathrm{\mu Jy}$] \\
  \hline
      L01 & 21 42 39.64 & -44 28 17.26 & $   81.4 \pm  5.3$ &   2.9 &         0.09  & $    4.1 \pm  0.3$ & $    4.7 \pm  0.6$ &                    & $   21.0 \pm  4.4$ \\
      L02 & 21 42 21.01 & -44 24 51.89 & $  911.0 \pm  5.4$ &   2.3 & \textbf{0.00} & $  190.8 \pm  0.7$ & $  135.1 \pm  0.7$ & $  139.1 \pm  3.1$ & $   92.7 \pm  4.6$ \\
      L03 & 21 42 47.54 & -44 27 29.19 & $   51.1 \pm  5.1$ &   5.3 &         0.19  & $    4.4 \pm  0.4$ & $    4.9 \pm  0.5$ & $   10.4 \pm  2.7$ &                    \\
      L04 & 21 42 22.07 & -44 28 09.02 & $  362.0 \pm  5.5$ &   3.3 & \textbf{0.02} & $   27.8 \pm  0.6$ & $   29.5 \pm  0.6$ & $   45.3 \pm  2.9$ & $   34.1 \pm  4.5$ \\
      L05 & 21 42 31.74 & -44 23 42.51 & $  230.0 \pm  5.6$ &   6.2 &         0.10  & $   15.3 \pm  1.2$ & $   14.9 \pm  1.9$ & $   22.7 \pm  3.5$ & $   22.7 \pm  4.4$ \\
      L06 & 21 42 34.97 & -44 27 08.70 & $  292.0 \pm  5.7$ &   3.4 & \textbf{0.03} & $   10.8 \pm  1.1$ & $   13.9 \pm  1.2$ & $   19.9 \pm  4.5$ & $   24.0 \pm  5.3$ \\
      L07 & 21 42 50.74 & -44 30 46.07 & $  175.0 \pm  6.6$ &   5.5 &         0.11  & $    2.4 \pm  0.6$ & $    3.3 \pm  0.7$ & $   10.2 \pm  3.1$ &                    \\
          & 21 42 51.49 & -44 30 43.18 & $ 1640.0 \pm  5.9$ &   2.6 & \textbf{0.00} & $  105.0 \pm  0.6$ & $  163.0 \pm  0.9$ & $  293.2 \pm  2.8$ & $  475.2 \pm  5.5$ \\
      L08 & 21 42 59.27 & -44 25 00.16 & $  355.0 \pm  5.2$ &   6.5 &         0.07  & $   17.3 \pm  0.4$ & $   22.9 \pm  0.6$ & $   32.0 \pm  2.6$ & $   21.0 \pm  4.5$ \\
      L09 &             &              &                    &       &               &                    &                    &                    &                    \\
      L10 &             &              &                    &       &               &                    &                    &                    &                    \\
      L11 & 21 42 44.73 & -44 23 28.13 & $  659.0 \pm  5.5$ &   2.9 & \textbf{0.01} & $   44.5 \pm  2.2$ & $   57.7 \pm  1.4$ & $   70.3 \pm  5.8$ & $   46.3 \pm  6.9$ \\
      L12 & 21 42 28.95 & -44 22 11.12 & $   99.4 \pm  5.7$ &   6.0 &         0.17  & $    5.0 \pm  0.5$ & $    7.1 \pm  0.6$ & $   11.2 \pm  3.0$ & $   15.7 \pm  4.7$ \\
      L13 & 21 43 03.06 & -44 25 12.13 & $  109.0 \pm  5.2$ &   8.0 &         0.19  & $    6.3 \pm  0.3$ & $    7.8 \pm  0.6$ &                    &                    \\
          & 21 43 02.64 & -44 25 03.15 & $  238.0 \pm  5.2$ &   3.4 & \textbf{0.04} & $   13.9 \pm  0.3$ & $   16.1 \pm  0.6$ & $   22.2 \pm  3.0$ & $   18.7 \pm  4.7$ \\
      L14 & 21 42 31.91 & -44 31 30.09 & $  155.0 \pm  5.1$ &   5.5 &         0.12  & $   32.5 \pm  0.4$ & $   21.6 \pm  0.5$ & $   22.9 \pm  2.6$ & $   23.3 \pm  4.2$ \\
      L15 & 21 42 49.30 & -44 30 29.73 & $  372.0 \pm  5.8$ &   2.8 & \textbf{0.02} & $   24.5 \pm  0.4$ & $   33.4 \pm  0.6$ & $   53.3 \pm  3.0$ & $   32.0 \pm  4.4$ \\
      L16 & 21 42 40.91 & -44 25 17.56 & $  143.0 \pm  5.6$ &   2.7 &         0.05  & $    9.9 \pm  0.4$ & $   11.6 \pm  0.7$ & $   14.1 \pm  3.0$ &                    \\
      L17 & 21 42 09.94 & -44 26 06.06 & $   59.3 \pm  5.7$ &   4.4 &         0.16  & $    4.4 \pm  0.3$ & $    6.0 \pm  0.6$ &                    &                    \\
      L18 & 21 42 33.58 & -44 22 09.80 & $  229.0 \pm  5.1$ &   7.9 &         0.13  & $    8.7 \pm  0.7$ & $   11.3 \pm  0.9$ & $   16.1 \pm  2.8$ &                    \\
          & 21 42 34.81 & -44 22 10.35 & $  180.0 \pm  5.4$ &   6.4 &         0.13  &                    &                    &                    &                    \\
      L19 & 21 42 16.07 & -44 28 49.50 & $  139.0 \pm  5.8$ &   2.3 & \textbf{0.04} & $    9.9 \pm  0.5$ & $   13.4 \pm  0.8$ & $   20.2 \pm  3.0$ & $   28.4 \pm  5.1$ \\
      L20 & 21 42 41.21 & -44 30 47.57 & $  113.0 \pm  5.4$ &   4.0 &         0.11  & $   33.5 \pm  0.4$ & $   31.0 \pm  0.5$ & $   21.0 \pm  2.7$ & $   92.4 \pm  4.3$ \\
          & 21 42 41.12 & -44 30 37.23 & $   62.8 \pm  5.4$ &   7.9 &         0.20  & $    8.1 \pm  0.5$ & $   12.7 \pm  0.6$ & $   23.6 \pm  3.1$ & $   29.6 \pm  4.8$ \\
          & 21 42 40.47 & -44 30 51.02 & $   78.1 \pm  5.4$ &   7.2 &         0.19  & $   71.0 \pm  0.7$ & $   45.6 \pm  0.8$ & $   40.9 \pm  2.7$ & $   26.6 \pm  4.7$ \\
      L21 &             &              &                    &       &               &                    &                    &                    &                    \\
      L22 &             &              &                    &       &               &                    &                    &                    &                    \\
 \hline
      L23 &             &              &                    &       &               &                    &                    &                    &                    \\
      L24 & 21 42 52.84 & -44 24 55.95 & $   94.6 \pm  5.9$ &   2.6 &         0.07  & $   31.5 \pm  0.4$ & $   21.4 \pm  0.5$ & $   29.6 \pm  3.3$ &                    \\
          & 21 42 52.85 & -44 24 56.16 & $   97.7 \pm  7.3$ &   2.0 & \textbf{0.05} &                    &                    &                    &                    \\
      L25 &             &              &                    &       &               &                    &                    &                    &                    \\
      L26 & 21 42 26.22 & -44 21 52.39 & $  172.0 \pm  5.9$ &   4.8 &         0.09  & $   17.1 \pm  0.4$ & $   22.3 \pm  0.6$ & $   41.9 \pm  3.0$ & $   23.7 \pm  5.0$ \\
      L27 & 21 43 11.73 & -44 26 16.94 & $   56.5 \pm  5.6$ &   2.0 &         0.07  & $   16.3 \pm  0.9$ & $   16.7 \pm  0.7$ & $   18.1 \pm  3.2$ & $   18.9 \pm  4.7$ \\
      L28 & 21 42 37.36 & -44 25 29.41 & $   79.4 \pm  5.6$ &   6.7 &         0.19  & $   19.7 \pm  0.4$ & $   17.9 \pm  0.5$ & $   21.0 \pm  2.9$ &                    \\
          & 21 42 37.98 & -44 25 35.86 & $   71.8 \pm  5.7$ &   6.5 &         0.19  &                    &                    &                    &                    \\
      L29 & 21 43 04.22 & -44 22 06.52 & $  206.0 \pm  5.4$ &   7.0 &         0.12  & $   56.2 \pm  0.6$ & $   66.8 \pm  0.6$ & $   65.6 \pm  3.0$ & $   47.1 \pm  4.3$ \\
      L30 &             &              &                    &       &               &                    &                    &                    &                    \\
      L31 & 21 42 31.05 & -44 26 44.61 & $  359.0 \pm  5.8$ &   2.4 & \textbf{0.01} & $   82.4 \pm  2.0$ & $   59.6 \pm  1.4$ & $   63.9 \pm  3.7$ & $   27.3 \pm  4.8$ \\
          & 21 42 30.73 & -44 26 43.02 & $  105.0 \pm  6.1$ &   5.3 &         0.15  &                    &                    &                    &                    \\
      L32 &             &              &                    &       &               &                    &                    &                    &                    \\
\end{tabular}
\end{minipage}
\end{table}

\end{document}